\documentclass[12pt,a4paper,oneside]{article}
\usepackage{ifpdf}
\usepackage{a4wide}
\usepackage{amsmath}
\usepackage{graphicx}
\usepackage{rotating}
\begin{document}
\setlength{\parskip}{0.45cm}
\setlength{\baselineskip}{0.75cm}



\begin{titlepage}
\setlength{\parskip}{0.25cm}
\setlength{\baselineskip}{0.25cm}
\begin{flushright}
DO-TH 07/06\\
\vspace{0.2cm}
August 2007
\end{flushright}
\vspace{1.0cm}
\begin{center}
\Large
{\bf Dynamical parton distributions of the nucleon}\\ 
{\bf and very small-x physics}
\vspace{1.5cm}

\large
M.~Gl\"uck, P.~Jimenez-Delgado, E.\ Reya\\
\vspace{1.0cm}

\normalsize
{\it Universit\"{a}t Dortmund, Institut f\"{u}r Physik}\\
{\it D-44221 Dortmund, Germany}

\vspace{1.5cm}
\end{center}

\begin{abstract}
\noindent Utilizing recent DIS measurements ($F_{2,\rm L}$) and data on
dilepton and high--$E_{\rm T}$ jet production we determine the dynamical
parton distributions of the nucleon generated radiatively from 
valence--like {\em positive} input distributions at optimally chosen
low resolution scales.  These are compared with `standard' distributions
generated from positive input distributions at some {\em fixed} and
higher resolution scale.
It is shown that up to the next to leading order 
NLO($\overline{\rm MS}$, DIS)
of perturbative QCD considered in this paper, the uncertainties of the 
dynamical distributions are, as expected, smaller than those of their 
standard counterparts.  This holds true in particular in the presently
unexplored extremely small--$x$ region relevant for evaluating ultrahigh
energy cross sections in astrophysical applications.  
It is noted that our new dynamical distributions are compatible, within
the presently determined uncertainties, with previously determined
dynamical parton distributions. 
\end{abstract}
\end{titlepage}


\section{Introduction}
The parton distributions of the nucleon are extracted from deep inelastic
scattering data by essentially two different approaches which differ in
their choice of the input distributions at some low scale $Q_0$. In
the common approach, e.g.\ \cite{ref1,ref2,ref3}, $Q_0$ is fixed at some
{\em arbitrarily chosen} $Q_0>1$ GeV and the corresponding input
distributions are unrestricted, allowing even for negative gluon
distributions \cite{ref3,ref4,ref5} in the small Bjorken-$x$ region, 
i.e.\ negative cross sections like $F_{\rm L}(x,Q^2)$.  
Alternatively \cite{ref6,ref7,ref8} the parton distributions at 
$Q$ \raisebox{-0.1cm}{$\stackrel{>}{\sim}$} 1 GeV are QCD radiatively
generated from {\em valence}--like {\em positive} input distributions
at an {\em optimally determined} $Q_0\equiv\mu <1$ GeV (where  
`valence--like' refers to $a_f>0$ for {\em all} input distributions
$xf(x,\mu^2)\propto  x^{a_f}(1-x)^{b_f}$).  This more restrictive ansatz implies,
of course, less uncertainties concerning the behavior of the parton
distributions in the small--$x$ region at $Q>\mu$ which is entirely due
to QCD dynamics.  In particular it provided unique (steep) predictions
\cite{ref6,ref9} for the experimentally unexplored region $x<10^{-2}$
which were subsequently first confirmed in \cite{ref10,ref11}.

This predictive power is especially important for investigations
concerning cross sections \cite{ref12} of ultrahigh energy particles
(neutrinos) produced via astrophysical acceleration processes, e.g.\
in active galactic nuclei, black holes or in the decays of very massive
particles (see, for example, \cite{ref13,ref14,ref15}).  Here one needs
a somewhat reliable knowledge of parton distributions at the weak scale
$Q^2=M_W^2$ down to $x\simeq 10^{-9}$ 
($x\simeq M_W^2/2m_N E_\nu$) at highest energies of
$E_\nu\simeq 10^{12}$ GeV which requires extrapolations into the yet
unmeasured small-$x$ region $x<10^{-3}$.  Furthermore this `radiative'
approach based on QCD dynamics is also useful for connecting 
nonperturbative models valid at $Q<1$ GeV (like chiral quark--soliton
models [16--20] and statistical parton models
[21--24]) with the actually measured distributions at
$Q>1$ GeV.

In the present paper we confront recent precision deep inelastic 
scattering (DIS) ep data, Drell--Yan dilepton and high--$p_{\rm T}$ jet data
with radiatively generated parton distributions arising from a 
valence--like positive input at $Q<1$ GeV, following and extending the
latest GRV98 analysis \cite{ref8}.  Moreover we study the dependence
and stability of the small-$x$ predictions, in particular of their
extrapolations down to $10^{-9}$, with respect to a different choice
of the factorization scheme ($\overline{\rm MS}$ versus DIS).  Furthermore
we compare these `dynamical' results with the ones obtained from the
common evolution approach being based on a non--valence--like input
at $Q_0>1$ GeV.  In particular we shall compare their associated
uncertainties.  As should be clear by now, it will turn out that these
uncertainties are indeed smaller for the radiatively generated parton
distributions, particularly in the small-$x$ region, due to their
valence--like input and the sizeably larger evolution distance
starting at $Q_0<1$ GeV. 

\section{Formalism}

The aforementioned analyses are undertaken at the next--to--leading
order (NLO) of perturbative QCD within the modified minimal substraction
($\overline{\rm MS}$) factorization and renormalization scheme. For
the radiative model we shall also present results as obtained within
the so--called DIS factorization scheme \cite{ref25,ref26}. Heavy quarks
($c, b, t$) will not be considered as partons, i.e.\ the number of
active flavors $n_f$ appearing in the splitting functions and the
corresponding Wilson coefficients will be fixed, $n_f=3$.  This defines
the so--called  `fixed--flavor number scheme' (FFNS).  As argued in
\cite{ref27}, it is nevertheless consistent and correct to utilize
the standard variable $n_f$ scheme for the $\beta$--function, and we
shall adopt this procedure in our evaluation of the running coupling
constant $\alpha_s(Q^2)$. Up to NLO, 
the strong coupling $a(Q^2)\equiv \alpha_s(Q^2)/4\pi$
evolves according to
\begin{equation}
da/d\ln Q^2 = -\beta_0 a^2-\beta_1 a^3
\end{equation}
where $\beta_0=11-2n_f/3$ and $\beta_1=102-38 n_f/3$.  Here we utilize
the exact numerical (iterative) solution for $a(Q^2)$ since it is
mandatory in the low $Q^2$ region \cite{ref8} relevant for the 
valence--like approach. The exact solution of (1) can be written
implicitly
\begin{equation}
\ln\frac{Q^2}{\tilde{\Lambda}^2} = \frac{1}{\beta_0 a(Q^2)} 
  -\frac{\beta_1}{\beta_0^2}\ln
    \left(\frac{1}{\beta_0 a(Q^2)} +\frac{\beta_1}{\beta_0^2}\right)\, .
\end{equation}
Since $\beta_{0,1}$ are not continuous for different $n_f$, the
continuity of $a(Q^2)$ requires to choose different values for the 
integration constant $\tilde{\Lambda}$ for different flavor numbers
$n_f$, $\tilde{\Lambda}^{(n_f)}$, which are fixed by the $a(Q^2)$ 
matchings at $Q=m_{c,b,t}$.  We have chosen
\begin{equation}
m_c = 1.3\,{\rm GeV},\quad\quad
  m_b = 4.2\,{\rm GeV},\quad\quad
    m_t = 175\,{\rm GeV}\, ,
\end{equation}
which turn out to be the optimal choices for all our subsequent LO
and NLO analyses of heavy quark production.  This exact solution
reduces to the common approximate  `asymptotic' solution
\begin{equation}
a(Q^2)\simeq \frac{1}{\beta_0\ln(Q^2/\Lambda^2)} -
     \frac{\beta_1}{\beta_0^3} \,\,  
       \frac{\ln\ln(Q^2/\Lambda^2)}{[\ln(Q^2/\Lambda^2)]^2}
\end{equation}
which turns out to be sufficiently accurate for $Q^2$
\raisebox{-0.1cm}{$\stackrel{>}{\sim}$} 2 GeV$^2$ and which, moreover,
is easier to use for practical applications of our results. The values
for $\Lambda^{(n_f)}$, as well as for $\tilde{\Lambda}^{(n_f)}$,
corresponding to our various LO and NLO global dynamical fits are 
given as follows: 
In LO, where $\beta_1 \equiv 0$ and $\tilde{\Lambda}_{\rm LO}^{(n_f)}
= \Lambda_{\rm LO}^{(n_f)},\, \Lambda_{\rm LO}^{(3,4,5,6)} = 210.3,\, 
181.8, \, 138.3, \, 70.1$ MeV;
in NLO($\overline{\rm MS}$) 
$\tilde{\Lambda}_{\overline{\rm MS}}^{(3,4,5,6)} = 302.8, \, 251.0,\, 
172.8,\, 70.0$ MeV and $\Lambda_{\overline{\rm MS}}^{(4,5,6)}=269.7,\, 
184.5,\, 72.9$ MeV;
in NLO(DIS) $\tilde{\Lambda}_{\rm DIS}^{(3,4,5,6)} = 288.5,\, 238.0,\,
162.9,\, 65.6$ MeV
while $\Lambda_{\rm DIS}^{(4,5,6)} = 255.9,\, 173.9,\, 68.3$ MeV.

Let us now turn to the update of our LO and NLO($\overline{\rm MS}$)
GRV98 distributions \cite{ref8} which consists of a fine tuning of the
valence--like input densities $xf(x,Q_0^2)$  as well as of the input
scale $Q_0\equiv\mu< 1$ GeV.  The non--singlet input densities 
$u_v,\, d_v,\, \Delta\equiv\bar{d}-\bar{u}$ and the valence--like
input distributions $\bar{d}+\bar{u},\, \bar{s}=s$ and $g$ in the
singlet sector 
are generically parametrized as
\begin{equation}
xf(x,Q_0^2) = N_f\, x^{a_f}(1-x)^{b_f}(1 + A_f\sqrt{x} +B_f x)
\end{equation}
subject to the constraints $\int_0^1 u_vdx = 2,
\int_0^1 d_vdx =1$ and
\begin{equation}
\int_0^1x[u_v+d_v+2(\bar{u}+\bar{d}+\bar{s}) +g]\, dx =1\, .
\end{equation}
Since the data sets we are using are insensitive to the specific choice
of the strange quark distributions, we continue to generate the
strange densities entirely radiatively \cite{ref8} starting from
$s(x,Q_0^2)=\bar{s}(x,Q_0^2)=0$ in the valence--like approach where
$Q_0<1$ GeV.  In the common standard approach where $Q_0>1$ GeV we
choose as usual 
$s(x,Q_0^2)=\bar{s}(x,Q_0^2)= [\bar{u}(x,Q_0^2)+\bar{d}(x,Q_0^2)]/4$.
Furthermore, since all our fits did not require the additional
polynomial in (5) for the gluon distribution, we have set 
$A_g=B_g=0$.  This left us with a total of 21 independent fit 
parameters, including $\alpha_s$.

These free parameters have been fixed using the following data sets.
The HERA ep measurements \cite{ref28,ref29} for $Q^2\geq 2$ GeV$^2$
for the   `reduced' DIS one--photon exchange cross section
$\sigma_r = F_2-(y^2/Y_+)F_{\rm L}$ together with the full neutral
current ($\gamma,\, Z^0)$ sector \cite{ref30}
\begin{equation}
\sigma_{r,\rm NC}^{e^{\pm}p}(x,Q^2) \equiv
 \left( \frac{2\pi\alpha^2 Y_+}{xQ^4}\right)^{-1} 
   \frac{d^2\sigma_{\rm NC}^{e^{\pm}p}}{dx\,dQ^2} =
     F_2^{\rm NC} -\frac{y^2}{Y_+}\, F_{\rm L}^{\rm NC} \mp
       \frac{Y_-}{Y_+}\, xF_3^{\rm NC}
\end{equation}
where $Y_{\pm} = 1\pm (1-y)^2$ and
\begin{eqnarray}
F_{2,{\rm L}}^{\rm NC} 
   & = & F_{2,{\rm L}} -v_e \kappa\, F_{2,{\rm L}}^{\gamma Z}
  + (v_e^2 + a_e^2)\kappa^2 F_{2,{\rm L}}^Z\nonumber\\
F_3^{\rm NC} 
  & = & -a_e\kappa\, F_3^{\gamma Z} + (2v_e a_e)\kappa^2\, F_3^Z
\end{eqnarray}
with $v_e= -\frac{1}{2} + 2\sin^2\theta_W$, $a_e= -\frac{1}{2},
\kappa^{-1} = 4\sin^2\theta_W\cos^2 \theta_W(Q^2+M_Z^2)/Q^2$
using $\sin^2\theta_W =0.2312$ and $M_Z=91.1876$ GeV.
Note that the structure functions in (7) refer to the radiatively
corrected ones as presented by the experimentalists. 
Furthermore, the well--known standard target mass corrections to
$F_2$ have been taken into account in the medium to large $x$--region
for $Q^2<100$ GeV$^2$.
 Since the 
experimental extraction of the usual (one--photon exchange) 
$F_2(x,Q^2)$ from $d^2\sigma/dx\, dQ^2$ is obviously (parton)
model dependent, we have chosen to work with the full NC framework
in order to avoid any further dependence on model assumption. 
However, it turned out that fitting just to $F_2(x,Q^2)$ gives 
very similar results.  Furthermore, since the longitudinal structure
function $F_{\rm L}$ contributes sizeably only for large 
inelasticity $y$, in most of the kinematic range we have $\sigma_r
\simeq F_2$.  In addition we have used the fixed target $F_2^p$ 
data of SLAC \cite{ref31}, BCDMS \cite{ref32}, E665 \cite{ref33}
and NMC \cite{ref34} subject to the standard cuts $Q^2 \geq 4$
GeV$^2$ and $W^2=Q^2(\frac{1}{x}-1) + m_p^2\geq 10$ GeV$^2$,
together with the structure function ratios $F_2^n/F_2^p$ of
BCDMS \cite{ref35}, E665 \cite{ref36} and NMC \cite{ref37}.
The data for heavy quark ($c,b$) production, being theoretically
described in the fixed--flavor number factorization scheme by the fully
predictive {\em fixed}--order (LO/NLO) perturbation theory to be
discussed below, are taken from \cite{ref38,ref39,ref40} for
$F_2^c$ and from \cite{ref40} for $F_2^b$.  Furthermore the 
Drell--Yan muon pair production data of E866/NuSea \cite{ref41}
for $d^2\sigma^{pN}/dx_F dM_{\mu^+\mu^-}$ with $N=p,d$ have
been used, as well as their asymmetry measurements \cite{ref42}
for $\sigma^{pd}/\sigma^{pp}$.  These data are instrumental in
fixing $\bar{d}-\bar{u}$ (or $\bar{d}/\bar{u}$). The relevant
LO/NLO differential Drell-Yan cross sections can be found in the 
Appendix of \cite{ref43}, except for eq.~(A8) which has to be
modified \cite{ref44,ref45} in order to conform with the usual
$\overline{\rm MS}$ convention for the number of gluon polarization
states $2(1-\varepsilon)$ in $4-2\varepsilon$ dimensions.  Finally,
the Tevatron high-$p_{\rm T}$ (or $E_{\rm T}$) inclusive jet data
of D0 \cite{ref46} and CDF \cite{ref47} have been used together
with the fastNLO \cite{ref48} package for calculating the relevant
cross sections at NLO. It should be mentioned that all these
data sets correspond to a total of of 1739 data points.

As already noted, the LO and NLO heavy quark contributions
$F_i^{c,b}$ are calculated in the FFNS and contribute to the total
structure functions as 
$F_i(x,Q^2)=F_i^{\rm light}+ F_i^{\rm heavy}$ where `light'
refers to the common $u,\, d,\, s$ (anti)quarks and gluon initiated
contributions \cite{ref44}, and $F_i^{\rm heavy} = F_i^c +F_i^b$.
Top quark contributions are negligible.  
The LO ${\cal{O}}(\alpha_s)$
contributions\footnote{It has become common to consider the
 ${\cal{O}}(\alpha_s)$ contributions to $F_{\rm L}^{\rm light}$ as
LO \cite{ref49}.  Therefore the 2--loop  ${\cal{O}}(\alpha_s^2)$
Wilson coefficients \cite{ref49} are required for 
$F_{\rm L}^{\rm light}$ at NLO. (Simplified parametrizations of
the relevant Wilson coefficients can be found in \cite{ref50}.)} to
$F_{2, \rm L}^{h}$, due to the subprocess $\gamma^*g\to h\bar{h}$,
have been summarized in \cite{ref7}, and the NLO  
${\cal{O}}(\alpha_s^2)$ ones are given in \cite{ref51,ref52}.  
These contributions are gluon $g(x,\mu_F^2)$ dominated
where the factorization scale should preferably be chosen
\cite{ref53} to be $\mu_F^2 = 4 m_h^2$.  As we shall see
the resulting predictions are in perfect agreement with all 
available DIS data on heavy quark production and are futhermore
perturbatively stable \cite{ref53}.  Even choosing a very large
scale like $\mu_F^2=4(Q^2+4m_c^2)$ leaves the NLO results 
essentially unchanged \cite{ref8,ref54} in particular at small-$x$.
This stability renders attempts to resum supposedly  `large
logarithms' ($\ln Q^2/m_h^2$) in heavy quark production cross
sections superfluous.  The heavy quark contributions to the 
remaining NC structure functions $F_{2, \rm L}^{\gamma \rm Z}$
and $F_{2, \rm L}^{\rm Z}$ in (5) are quantitatively negligible.
Nevertheless, they can be simply obtained from the 
$\gamma$--exchange contributions $F_{2,L}^h$ by substitutions
like $e_q^2\to 2e_q\, v_q$ and $e_q^2\to v_q^2+a_q^2$, respectively,
where $v_q=\pm\frac{1}{2}-2e_q\sin^2\theta_W$ and 
$a_q=\pm\frac{1}{2}$ with $\pm$ referring to an up-- or
down--type quark.  The heavy quark contributions to 
$F_3^{\gamma\rm Z,Z}$ vanish in LO \cite{ref55,ref56}, and are
negligibly small in NLO since effectively $F_3^h\sim h-\bar{h}$
at the relevant large values of $Q^2$.

In order to test the dependence of our results on the specific
choice of the factorization scheme in NLO, other than the 
commonly used $\overline{\rm MS}$ scheme, we also perform our
NLO analysis using the deep inelastic scattering (DIS) factorization
scheme \cite{ref25,ref26}.  Here the $\overline{\rm MS}$ Wilson
coeffients are absorbed into the parton distributions, or more
precisely into their evolutions, i.e., into the splitting functions.
This transformation to the DIS scheme in NLO is achieved via
\cite{ref44}
\begin{equation}
P_{\rm NS}^{(1)}\to P_{\rm NS,\,DIS}^{(1)} = P_{\rm NS}^{(1)}
   +\beta_0\Delta C_{2,\rm NS}^{(1)}
\end{equation}
\begin{equation}
\hat{P}^{(1)}\to\hat{P}_{\rm DIS}^{(1)} = \hat{P}^{(1)} +
 \beta_0\Delta \hat{C}_2^{(1)} -
\left[\Delta \hat{C}_2^{(1)}\otimes \hat{P}^{(0)}-\hat{P}^{(0)}
 \otimes \Delta\hat{C}_2^{(1)}\right]
\end{equation}
where
\begin{equation}
\Delta C_{2,\rm NS}^{(1)} = - C_{2,\rm NS}^{(1)}\quad, \quad\quad
\Delta \hat{C}_2^{(1)} = -
\left( 
\renewcommand{\arraystretch}{1.5}\begin{array}{ccc}
C_{2,q}^{(1)} & , &  C_{2,g}^{(1)}\\
-C_{2,q}^{(1)} & , & -C_{2,g}^{(1)} \end{array} \right)\, .
\end{equation}
The light $u,\, d,\, s$ quark contributions to $F_2^p$, for example,  
in the NLO(DIS) factorization scheme now simply become
\begin{equation}
F_2^{\rm light}(x,Q^2) = x\sum_{q=u,d,s} e_q^2
   \left[q(x,Q^2)+\bar{q}(x,Q^2)\right]_{\rm DIS}\, .
\end{equation}
The quantitative difference between the NLO($\overline{\rm MS}$)
and NLO(DIS) results will turn out to be rather small. 
Having obtained the parton distributions 
\raisebox{-0.05cm}{$\stackrel{(-)}{q}$}$\!(x,Q^2)_{\rm DIS}$ 
and $g(x,Q^2)_{\rm DIS}$
from an explicit NLO analysis of $F_2(x,Q^2)$ in the DIS factorization
scheme, one can transform them to the $\overline{\rm MS}$ scheme
via (see \cite{ref7}, for example)
\begin{eqnarray}
\text{\raisebox{-0.055cm}{$\stackrel{(-)}{q}$}}\!(x,Q^2) & = & 
\text{\raisebox{-0.055cm}{$\stackrel{(-)}{q}$}}\!(x,Q^2)_{\rm DIS}
 -a\left[C_{2,q}^{(1)}\otimes\!\!
\text{\raisebox{-0.055cm}{$\stackrel{(-)}{q}$}}\!_{\rm DIS} +
   \frac{1}{2f}\, C_{2,g}^{(1)}\otimes g_{\rm DIS}\right]\!
     (x,Q^2)+{\cal{O}}(a^2)\\
g(x,Q^2) & = & g(x,Q^2)_{\rm DIS}
 +a\left[ C_{2,q}^{(1)}\otimes \Sigma_{\rm DIS} + C_{2,g}^{(1)}
    \otimes g_{\rm DIS}\right]\!
      (x,Q^2) +{\cal{O}}(a^2)
\end{eqnarray}
where
\begin{eqnarray}
C_{2,q}^{(1)}(z)&  = & 2\frac{4}{3}
 \left[\frac{1+z^2}{1-z} \left(\ln \frac{1-z}{z} -\frac{3}{4}\right)
  +\frac{1}{4}\, (9+5z)\right]_+\\
C_{2,g}^{(1)}(z) & = & 4n_f\frac{1}{2} 
 \left[(z^2+(1-z)^2)\ln\frac{1-z}{z} -1+8z(1-z)\right]
\end{eqnarray}
with $n_f=3$.  This transformation to the $\overline{\rm MS}$ scheme then
allows also for a consistent NLO analysis of heavy quark and Drell--Yan dimuon
production processes in the DIS scheme, using their well known theoretical
$\overline{\rm MS}$ expressions, as well as for a consistent comparison
of our DIS results with the ones obtained in the $\overline{\rm MS}$
factorization scheme.
\vspace{0.5cm}
 
\noindent{\large\bf{2a. Estimates of uncertainties}}

Our evaluation of the parton distribution uncertainties is based on the 
Hessian method with the Hessian matrix defined via
\begin{equation}
\Delta\chi^2 =\chi^2-\chi_0^2 = \frac{1}{2}\sum_{i,j,=1}^d H_{ij}
   (a_i -a_i^0) (a_j-a_j^0)
\end{equation}
where $\chi_0^2$ is the value of the minimal $\chi^2$ characterized by
the free fit parameters $a_i^0$.  In our fits we have $d=21$ and $\chi^2$
is calculated by adding the total systematic and statistical
experimental errors in quadrature.  The uncertainties $\Delta a_i=
a_i -a_i^0$ are constrained by 
\begin{equation}
\Delta\chi^2\leq T^2
\end{equation}
where the tolerance parameter $T$ was chosen to be \cite{ref57}
\begin{equation}
T^2 = T_{1\sigma}^2 = \sqrt{2N} / (1.65)^2\, ,
\end{equation}
i.e.\ $T\simeq 4.7$ since $N=1739$ is the total number of data points
considered in our global fits.  The inversion involved in evaluating
$\Delta a_i$ in (17), subject to the constraint (18), was performed
with the help of the normalized eigenvectors \cite{ref58} of $H_{ij}$
whose iterative calculation followed \cite{ref59}.  The calculation of
all the uncertainties presented in our paper was performed according 
to the master equation (24) of 
\cite{ref58} whose particular implication
for $\Delta a_i$ is specified in eq.~(30) of \cite{ref58}.  Our choice for
the displacement distance $t$ entering these latter equations was $t=T$,
an assumption made in most subsequent publications and analyses.
(When comparing our uncertainty results with the ones of CTEQ 
\cite{ref2,ref58} where $T=10$ has been assumed, we rescale these CTEQ
uncertainties according to $\Delta a_i\to 0.47 \Delta a_i$ in order to
comply with our $T=4.7$ in (19).)

As suggested in \cite{ref2}, we included in our final error analysis only
those parameters that are actually sensitive to the input data set chosen,
i.e.\ those parameters which are not close to `flat' directions in the 
overall parameter space. With current data, and our functional form (5),
13 such parameters, including $\alpha_s$, are identified and are included
in our final error analysis; the remaining ill--determined eight
polynomical parameters $A_f$ and $B_f$, with uncertainties of more than
50\%, were held fixed.

\section{Quantitative results and very small-$x$ predictions}

A representative comparison of our dynamical LO and 
NLO($\overline{\rm MS}$)
results with the relevant HERA(H1, ZEUS) data on the proton structure
function $F_2^p(x,Q^2)$ is presented in Figs.\ 1 and 2.  Due to our
valence--like input, the small--$x$ results
($x$ \raisebox{-0.1cm}{$\stackrel{<}{\sim}$}  $10^{-2}$) are 
{\em predictions} being entirely generated by the QCD $Q^2$--evolutions.
This is in contrast to a  `standard fit' where the gluon and sea input
distributions in (5) do {\em not} vanish as $x\to 0$ 
\mbox{($a_{g,\bar{q}}$
\raisebox{-0.1cm}{$\stackrel{<}{\sim}$} 0)} at $Q_0^2 =2$ GeV$^2$.
For comparison we have also performed such a standard fit shown by
the dashed--dotted curves in Figs.~1 and 2.  In both cases the data
in Figs.~1 and 2 are well described throughout the whole 
medium-- to small--$x$ region for 
$Q^2$ \raisebox{-0.1cm}{$\stackrel{>}{\sim}$} 
2 GeV$^2$ and thus perturbative QCD is here fully operative.  At 
$Q^2<2$ GeV$^2$ the theoretical results fall below the data in the 
very small--$x$ region; this is not unexpected for perturbative leading 
twist--2 results, since nonperturbative (higher twist) contributions to 
$F_2(x,Q^2)$ will eventually become relevant, even dominant, for 
decreasing values of $Q^2$.  It should be emphasized that all of our 
valence--like input distributions at $\mu_{\rm LO}^2=0.3$ GeV$^2$ and 
$\mu_{\rm NLO}^2=0.5$ GeV$^2$ as well as the ones for the `standard fit' 
at $Q_0^2=2$ GeV$^2$ are manifestly {\em positive}.  This is in contrast 
to negative gluon distributions in the small--$x$ region observed in 
other standard fits \cite{ref3,ref4,ref5}. Furthermore the more 
restrictive ansatz of the valence--like input distributions at 
small--$x$ as well as the sizeably larger evolution distance (starting
at $Q_0<1$ GeV) imply {\em smaller} uncertainties concerning the 
behavior of structure functions in the small--$x$ region than the 
corresponding results obtained from the common   `standard fits',
in particular as $Q^2$ increases.  This is illustrated in Fig.~3
for the NLO($\overline{\rm MS}$) results where the error bands 
correspond to a $1\sigma$ uncertainty.  Since our valence--like
sea input has a rather small power of $x$, i.e.\ vanishes only slowly
as $x\to 0$, the uncertainties of the sea dominated $F_2(x,Q^2)$
turn out to be not too different from the standard fit where the 
sea increases as $x\to 0$ (negative power of $x$) already at the 
input scale $Q_0^2 = 2$ GeV$^2$.  Notice that the uncertainties 
generally decrease as $Q^2$ increases due to the QCD $Q^2$--evolutions
\cite{ref57,ref58}.

Our NLO($\overline{\rm MS}$) valence--like input distributions at
$Q_0^2 = \mu_{\rm NLO}^2 = 0.5$ GeV$^2$ together with their $1\sigma$
uncertainties are shown in Fig.~4.  They have been parametrized 
according to (5) with the parameters given in Table 1.\footnote{It
should be mentioned that there is a correlation between the (chosen)
value of $\alpha_s(M_{\rm Z}^2)$ and the resulting values of the
valence--like input scales $\mu_{\rm LO,NLO}$ which increase with
$\alpha_s(M_{\rm Z}^2)$.  Since we did not want to fix
$\alpha_s(M_{\rm Z}^2)$ at the LEP value of 0.118, we performed fits
for various fixed values of $\mu$ by imposing a valence--like input
structure $(a_{g,\bar{u}+\bar{d}}>0)$ and keeping $\alpha_s$ as a
free fit parameter.  Then we fixed the best choice for $\mu\, 
(\mu_{\rm LO}^2 = 0.3$ GeV$^2$, $\mu_{\rm NLO}^2 = 0.5$ GeV$^2$)
and performed the final precision fits and error analyses.}
For comparison the GRV98 input \cite{ref8} is displayed as well,
which turns out  to be very similar except for the gluon which peaks at a 
slightly larger value of $x$.  However, such differences are merely
within a $2\sigma$ band of our new results.  The valence--like
gluon input at low $Q^2<1$ GeV$^2$ in Fig.~4 implies a far stronger
constrained gluon distribution at larger values of $Q^2$ as compared
to a gluon density obtained from a `standard fit' with a 
conventional non--valence--like input at $Q^2>1$ GeV$^2$ as can be
seen in Fig.~5.  As already mentioned, this is in contrast to the 
sea distribution $\bar{u}+\bar{d}$ in Fig.~5 where the valence--like
sea input in Fig.~4 vanishes very slowly as $x\to 0$ and thus is
similarly increasing with decreasing $x$ down to $x\simeq 0.01$ as
the sea input obtained by a standard fit.  Therefore the $1\sigma$
uncertainty band of our dynamically predicted sea distributions at
larger values of $Q^2$ in Fig.~5 is only marginally smaller than
the corresponding one of the standard fit.  The relevant input
parameters of our  `standard fit' can be found in Table 2.  As 
expected for the dynamical fit, starting from a low input scale
with valence--like distributions, $\alpha_s(M_{\rm Z}^2)$ in Table 1
is somewhat stronger constrained due to the larger evolution 
distance than the corresponding result of the standard 
NLO($\overline{\rm MS}$) fit in Table 2.  Keeping in mind that our
stated errors always refer to $1\sigma$ uncertainties, our
standard fit error of 0.0021 for $\alpha_s(M_{\rm Z}^2)$ in 
Table 2 is compatible with the $2\sigma$ uncertainty also stated 
in the literature (see, e.g., \cite{ref2} and the discussion in
\cite{ref3}). It should be furthermore mentioned that our 
NLO($\overline{\rm MS}$) results for $\alpha_s(M_Z^2)$ in Table 1
and 2 are, within about a $1\sigma$ uncertainty, also compatible
with the ones obtained from analyzing only DIS structure functions
(for a recent summary, see \cite{ref60}).

At this point it should be mentioned that the standard CTEQ \cite{ref2}
fit resulted, very surprisingly, in a {\em{valence}}--like input
gluon distribution at a scale as large as $Q_0^2=m_c^2\simeq 1.7$
GeV$^2$.  Thus this CTEQ6 gluon distribution is expected to be 
similarly tightly constrained at $Q^2>Q_0^2$ as our dynamical results
starting from a valence--like input at $Q_0^2 = 0.5$ GeV$^2$.  That
this is indeed the case is illustrated in Fig.~6 where the $1\sigma$
uncertainties of the CTEQ6 gluon \cite{ref2} are similar in size to
our dynamical results, whereas a common  `standard fit' (being based
on an increasing input distribution as $x\to 0$) results in a 
sizeably larger uncertainty.  The situation is different for the sea 
distribution in the small--$x$ region;  here the CTEQ6 input at
$Q_0^2=m_c^2$ increases at $x\to 0$ as expected, and thus the 
$1\sigma$ uncertainty is comparable to our  `standard fit' result
as shown in Fig.~7 -- both being larger than the $1\sigma$ uncertainty
obtained from our dynamical fit based on valence--like inputs at
$Q_0^2 = 0.5$ GeV$^2$.

The heavy charm and bottom quark contributions to $F_2$ at LO and 
NLO($\overline{\rm MS}$) are compared with recent HERA data in 
Figs.~8 and 9.  The impressive agreement with present measurements
for $F_2^c$ and $F_2^b$ illustrates that the $n_f=3$ FFNS is 
entirely reliable.  As already discussed in Sec.~2, the NLO 
results are rather insensitive to the chosen factorization scale
$\mu_F$ ($\mu_F^2 = 4m_h^2$ or $\mu_F^2 = Q^2 + 4m_h^2$).  
Again the $1\sigma$ uncertainties of these dynamical
predictions are distinctly smaller than the ones implied by the 
standard fit.  It should be furthermore reemphasized that within the
FFNS heavy quarks ($h=c,b,t$) are always produced as {\em final}
states in fixed order perturbation theory via hard production 
processes initiated by the light partons of the nucleon ($u,d,s$
quarks and the gluon $g$).  The perturbative stability of heavy
quark production \cite{ref53} as well as the agreement with experiment
in Figs.~8 and 9 even at $Q^2\gg m_h^2$ indicate that there is 
{\em no} need to resum supposedly  `large logarithms' 
($\ln Q^2/m_h^2$), which is of course in contrast to genuine 
collinear logarithms appearing in light (massless) quark and gluon
hard scattering processes.  Therefore only the $n_f=3$ light $u,d,s$
quark flavors and gluons constitute the   `intrinsic' genuine
partons of the proton and the heavy $c,b,t$ quark flavors should 
not be included in the parton structure of the nucleon, not even
at $Q^2\gg m_h^2$ \cite{ref53}. 
However, somewhat dissenting views were recently summarized in
\cite{ref61}.

The measurements of Drell--Yan dilepton production in $pp$
and $pd$ collisions \cite{ref41,ref42} are instrumental in fixing
$\Delta = \bar{d}-\bar{u}$ (or $\bar{d}/\bar{u}$) \cite{ref62}.
In Fig.~10 we display our dynamical NLO($\overline{\rm MS}$) result
for $\sigma^{pd}/2\sigma^{pp}$ together with the $\pm 1\sigma$ 
uncertainty band as well as the previous GRV98 result which agree 
in the statistically relevant $x$--region, with $x_2$ referring to
the average fractional momentum of the target partons.  Note that
$\sigma^{pN}\equiv d^2\sigma^{pN}/dx_F dM_{\mu^+\mu^-}^2$ with
$x_F = x_1-x_2$.  In LO 
$\sigma^{pN}\sim \sum_{u,d,s}e_q^2\left[q(x_1)\bar{q}(x_2)+q(x_2)
\bar{q}(x_1)\right]$
where $x_1$ and $x_2$ refer to the Bjorken--$x$ of the quarks in the
beam ($p$) and nucleon target ($N$), respectively.  Experimentally
$x_F>0$ ($x_1>x_2$) and consequently the Drell--Yan cross
section is dominated by the annihilation of a beam quark with a
target antiquark.  For $x_1\gg x_2$ one obtains 
$\sigma^{pd}/2\sigma^{pp}\simeq\frac{1}{2}\left[1+\bar{d}(x_2)/
 \bar{u}(x_2)\right]$
at a scale $Q^2\equiv M_{\mu^+\mu^-}^2$ in $\bar{q}(x_2,\, Q^2)$.

Finally the $p\bar{p}$ Tevatron high--$p_{\rm T}$ (or $E_{\rm T}$)
inclusive jet data \cite{ref46,ref47} are compared in Fig.~11 with
our dynamical LO and NLO($\overline{\rm MS}$) results, as well as 
with the ones of CTEQ6 \cite{ref2}.  The small $1\sigma$ error
bands are almost invisible on the huge logarithmic scale used.
Our NLO result almost coincides with the one of CTEQ.  There is a 
clear improvement at NLO as compared to LO which falls slightly
below the data at $p_{\rm T}$ \raisebox{-0.1cm}{$\stackrel{<}{\sim}$}
300 GeV.  Nevertheless the LO high--$p_{\rm T}$ fit corresponds to
$\chi^2/\rm dof\simeq 1$ which is only twice as large as at NLO.

As discussed in Sec.~2 we have explicitly used for our analysis the
experimentally directly measured   `reduced' DIS cross sections
(7) which, for not too large values of $Q^2$, are dominated by the
one--photon exchange cross section 
$\sigma_r = F_2-(y^2/Y_+)F_{\rm L}$ where $y=Q^2/xs$.
The importance of using this quantity has recently been emphasized
\cite{ref63} since the effect of $F_{\rm L}$ becomes increasingly
relevant as $x$ decreases at a given $Q^2$ where $y$ increases.  This
is seen in the data as a flattening of the growth of $\sigma_r(x,Q^2)$
as $x$ decreases to very small values, at fixed $Q^2$, leading
eventually to a turnover (cf.~Fig.~12).  At the lower values of $Q^2$
in Fig.~12 it was not possible in \cite{ref63} to reproduce this 
turnover at NLO.  This was mainly due to the negative longitudinal
cross section (negative $F_{\rm L}(x,Q^2)$) encountered in 
\cite{ref63}. Since all of our cross sections and structure functions
are manifestly positive throughout the whole kinematic region
considered, our dynamical NLO($\overline{\rm MS}$) results in 
Fig.~12 are in good agreement with all small--$x$ HERA measurements
\cite{ref28,ref29}.  For completeness we compare in Fig.~13 our
dynamical (leading twist) NLO($\overline{\rm MS}$) predictions for
$F_{\rm L}(x,Q^2)$ with a representative selection of (partly
prelinimary) H1 data \cite{ref28,ref64} at fixed $W\simeq 276$ GeV. The
standard fit result with its sizeably larger $\pm 1\sigma$ error
band is, for comparison, shown as well.  Our NLO results for 
$F_{\rm L}$, being gluon dominated in the small--$x$ region, are
in full agreement with present measurements, which is in contrast
to expectations \cite{ref3,ref63} based on negative parton
distributions and structure functions at small $x$.  To illustrate
the manifest positive definiteness of our dynamically generated
structure functions we show $F_{\rm L}(x,Q^2)$ in Fig.~13 down to
$Q^2 = 1$ GeV$^2$ although a leading twist--2 prediction should
not be confronted with data below, say, 2 GeV$^2$.

As our parameter--free small--$x$ predictions for parton distributions
at $x<10^{-2}$ are entirely of QCD--dynamical origin and depend 
rather little on the detailed input parameters at $x$
\raisebox{-0.1cm}{$\stackrel{>}{\sim}$} $10^{-2}$, it is interesting to
study these predictions in kinematic regions not accessible by 
present DIS experiments.  Of particular interest are, as already
emphasized in the Introduction, calculations of weak
$\stackrel{(-)}{\nu}\!\!\!N$ cross sections of ultrahigh energy cosmic
neutrinos [12, 15, 65--69] which afford a (reliable)
knowledge of parton distributions at the weak scale $Q^2=M_W^2$
down to $x\simeq 10^{-9}$ for highest energies $E_{\nu}\simeq 10^{12}$
GeV.  This requires extrapolations into the unmeasured small--$x$
region $x<10^{-3}$.  Since $F_2^p(x,Q^2)$ is, in the very small--$x$
region, dominated by $\bar{u}(x,Q^2)$ and $\bar{d}(x,Q^2)$, as are
the CC neutrino--(isoscalar) nucleon cross sections, the $F_2^p$
structure function can be utilized for estimating the magnitude of 
uncertainties of the predictions in the extreme small--$x$ region
which are shown in Fig.~14.  At $Q^2=M_W^2$ our dynamical NLO
predictions correspond to a $\pm 1\sigma$ uncertainty of about $\pm$7\%
at $x=10^{-9}$ whereas the uncertainty of the extrapolation of a 
standard fit is about twice as large.  At smaller scales the 
uncertainties obviously increase as illustrated in Fig.~14 at 
$Q^2=100$ GeV$^2$.  Taking into account previous extrapolation 
ambiguities \cite{ref8}, one can conclude \cite{ref12} that the 
dynamically predicted small--$x$ parton distributions allow
neutrino--nucleon cross sections to be calculable with an accuracy
of about 10\% at highest cosmic neutrino energies.  It should be
mentioned that an ad hoc fixed power law of $x$ extrapolation of the
standard CTEQ6.5 structure functions \cite{ref70} to $x=10^{-8}$
at $Q^2=M_W^2$ \cite{ref71} lies, accidentally, only about 10\%
below our dynamical NLO prediction in Fig.~14.  On the other hand,
an alternative parametrization \cite{ref71} of present HERA(ZEUS)
data which is not QCD oriented but based on analyticity and 
unitarity gives, when extrapolated to $x = 10^{-8}$, a factor of
about 6 smaller a value for $F_2^p(10^{-8},\, M_W^2)$ than our
prediction in Fig.~14.  Since the perturbative dynamical QCD
{\em pre}dictions for the small--$x$ behavior of structure functions
down to $x=10^{-5}$ proved to be in agreement with later HERA
measurements as discussed in the Introduction, it is hard to 
imagine that perturbative QCD dynamics and evolutions should become
entirely inappropriate at $x=10^{-8}$ to $10^{-9}$ at even much
larger scales.

In order to test the dependence of our results on the specific
choice of the factorization scheme in NLO, we have repeated our 
dynamical analysis in the DIS factorization scheme as outlined in
Sec.~2.  Since the (light) parton distributions in the DIS scheme
are defined via the $F_2$ structure function in (12), it is not
very surprising that, in contrast to the DIS parton distributions
themselves \cite{ref72}, the results for physical observables directly
related to DIS structure functions are very similar.  Indeed the
results in the DIS factorization scheme for $F_2^p(x,Q^2)$ in 
Figs.~1 to 3 and 14 are practically indistinguishable from the ones
in the $\overline{\rm MS}$ scheme, as are the results for 
$F_2^{c,b}(x,Q^2)$ in Figs.~8 and 9 and the ones for 
$\sigma_r(x,Q^2)$ and $F_{\rm L}(x,Q^2)$ in Figs.~12 and 13,
respectively.  Differences become visible only for processes which
are not directly related to DIS structure functions such as 
Drell--Yan cross sections but there the differences lie well within
the $\pm 1\sigma$ uncertainty of the NLO($\overline{\rm MS}$) 
results as illustrated in Fig.~10 for the DY--asymmetry.  The 
parameters for the input parton distributions in the DIS scheme
and the corresponding $\alpha_s(M_Z^2)$ can be found in Table 1.
It should be furthermore emphasized that our results and predictions
are also stable to within less than about $20\%$ when compared to
{\em previous} analyses and fits.  This is illustrated in Fig.~15
for our present dynamical NLO($\overline{\rm MS}$) results when
compared with our previous GRV98 results \cite{ref8}.  The situation
is similar for more recent and previous standard CTEQ and MRST
parton distributions for their relevant ranges in $x$, and holds
also by comparing CTEQ and MRST distributions with each other 
\cite{ref2,ref3,ref57,ref70,ref73,ref74}.  
It should be emphasized that heavy quark mass effects have always
been fully taken into account in our previous \cite{ref7,ref8}
and present analyses.  This is in contrast to the previous CTEQ6
analysis \cite{ref2} where charm has been treated in the zero--mass
approximation.  The recent inclusion of finite charm mass effects
in CTEQ6.5 \cite{ref70} reduces the charm contribution to $F_2(x,Q^2)$
which is compensated by larger $u=u_v+\bar{u}$ and $d=d_v+\bar{d}$
distributions at small $x$ as compared to CTEQ6 \cite{ref2}.
That such an  `enhancement' has always been present in our dynamical
$u$ and $d$ distributions is illustrated in Fig.~16, since our 
present and previous (cf.~Fig.~15) distributions differ very little
from the CTEQ6.5 ones.  Therefore our predicted hadronic 
$W^{\pm}/Z^0$ production cross sections, for example, at Tevatron
and LHC are similar to the  `enhanced' ones observed in \cite{ref70}.

Of course more recent parton distributions have a higher precision
due to the higher statistics of the data, but we have not experienced
essential qualitative and quantitative changes during the past decade.
It is reassuring to see that our knowledge of the fundamental partonic
structure of matter has essentially remained unchanged over the past
years.

\section{Summary and Conclusions}

Utilizing recent DIS measurements and data on Drell--Yan dilepton
and high--$E_{\rm T}$ inclusive jet production, we have redone a
previous \cite{ref8} global fit for the dynamical parton distributions
of the nucleon in the LO and NLO of perturbative QCD.  The small--$x$
($x$ \raisebox{-0.1cm}{$\stackrel{<}{\sim}$} $10^{-2}$)
structure of dynamical parton distributions is generated entirely
radiatively from {\em valence--like}, manifestly {\em positive},
input distributions at an optimally chosen input scale $Q_0<1$ GeV.
The NLO results are stable with respect to a different choice of the 
factorization scheme ($\overline{\rm MS}$ versus DIS).  The 
predictions for the longitudinal structure function 
$F_{\rm L}(x,Q^2)$ at small $x$, for example, are positive throughout
the whole kinematic region considered, in agreement with (partly
preliminary) data.  We have augmented our analyses with an 
appropriate uncertainty analysis and found that the newly determined
dynamical distributions are compatible with the former 
\cite{ref7,ref8} ones, where heavy quark mass effects have always
been fully taken into account.  The stability of these results 
guarantees a reliable calculation of cross sections for, e.g., 
heavy quark, $W^{\pm}$, $Z^0$, and high--$p_{\rm T}$ jet production
at hadron colliders like Tevatron and in particular LHC.

Our dynamical distributions have also been compared with conventional
(`standard') ones obtained from {\em non}--valence--like positive
input distributions at some arbitrarily chosen higher input scale
$Q_0 >1$ GeV.  For this purpose we have performed a `standard fit'
as well, assuming $Q_0^2=2$ GeV$^2$.  The uncertainties of these
latter distributions are, as expected, larger, in particular in the 
present (experimentally) unexplored extremely small--$x$ region
relevant for evaluating ultrahigh energy neutrino--nucleon cross
sections in astrophysical applications.  Here we provide 
predictions down to $x\simeq 10^{-9}$ at the weak scale
$Q^2=M_W^2$ as required
\cite{ref12,ref65,ref66} for highest cosmic neutrino energies of
$10^{12}$ GeV.
These predictions are strongly constrained within the dynamical
parton model and are entirely of QCD-dynamical origin in the very
small--$x$ region.  Furthermore, as mentioned in the Introduction,
previous predictions \cite{ref6,ref9} for the small--$x$ region
based on the dynamical parton model and the data available at 
that time were subsequently confirmed \cite{ref10,ref11} at HERA.
The presently available very precise small--$x$ data 
\cite{ref28,ref29} utilized here allows us to be quite confident 
about the reliability of our improved small--$x$ predictions 
within the framework of the successful dynamical parton model.

A FORTRAN package (grid) containing our new dynamical LO, 
NLO($\overline{\rm MS}$), and NLO(DIS) parton densities, the light
($u,d,s;g)\,\, F_2^{\rm light}(x,Q^2)$ as well as $F_2^{c,b}(x,Q^2)$,
calculated in the fixed order FFNS, can be obtained by electronic
mail or on request.  The NLO($\overline{\rm MS}$) uncertainty
estimates will be also included.
\vspace{1.0cm}

\noindent{\underline{\bf Acknowledgements}}

\noindent  We thank J.~Bl\"umlein for helpful discussions.
This work has been supported in part by the `Bundesministerium 
f\"ur Bildung und Forschung', Berlin/Bonn.

\newpage
\pagestyle{empty}
\begin{sidewaystable}[th]
\renewcommand{\arraystretch}{1.8} 
\vskip 14pt
\scriptsize
\centering
\begin{center}
\begin{tabular}{|c||c|c|c|c|c||c|c|c|c|c||c|c|c|c|c|}
\hline
&\multicolumn{5}{|c||}{NLO ($\overline{\rm MS}$)}  & 
 \multicolumn{5}{|c||}{NLO (DIS)} &  
 \multicolumn{5}{|c|}{LO}\\
\hline
& $u_v$ & $d_v$ & $\bar{d} - \bar{u}$ & $\bar{u} + \bar{d}$ & $g$ &
  $u_v$ & $d_v$ & $\bar{d} - \bar{u}$ & $\bar{u} + \bar{d}$ & $g$ &
  $u_v$ & $d_v$ & $\bar{d} - \bar{u}$ & $\bar{u} + \bar{d}$ & $g$ \\
\hline
N & 1.2757 & 0.7893 & 4.0918 & 0.8627 & 3.1367 & 
    0.4341 & 0.1766 & 8.5986 & 0.9348 & 19.447 & 
    3.3434 & 0.3016 & 4.6430 & 0.6333 & 19.5921 \\ 

a & 0.4960 & 0.5165 & 1.5483 & 0.1450 & 0.5168 & 
    0.3069 & 0.2514 & 1.4181 & 0.1516 & 0.9146 & 
    0.6135 & 0.3473 & 1.4409 & 0.0224 & 1.3902 \\ 

b & 3.4525 & 4.6006 & 16.854 & 9.6252 & 2.7961 &  
    2.3124 & 3.3688 & 15.224 & 6.5321 & 6.6235 & 
    3.1866 & 3.6160 & 12.870 & 8.0034 & 4.5219 \\ 

A & -2.0704 & -1.8488 & -2.7767 & -1.7699 & - & 
     0.8040 & -0.4417 & -6.2906 & -1.2910 & - & 
    -3.3631 & -0.7803 & -2.8689 & -2.0277 & - \\ 

B & 13.225 & 14.179 & 24.257 & 7.9169 & - & 
    12.163 & 23.866 & 16.243 & 1.6333 & - & 
    7.6775 & 18.572 & 9.3879 & 6.5419 & - \\ 

\hline
$\chi^2/{\rm dof}$ & \multicolumn{5}{|c||}{1.061 (1.039)} & 
                     \multicolumn{5}{|c||}{1.073 (1.026)} & 
                     \multicolumn{5}{|c|}{1.295 (1.213)}  \\ 
\hline
$\alpha_s(M_Z^2)$ & \multicolumn{5}{|c||}{0.1145 $\pm$ 0.0018}  & 
                    \multicolumn{5}{|c||}{0.1135 $\pm$ 0.0019}  & 
                    \multicolumn{5}{|c|}{0.1263 $\pm$ 0.0015 }   \\ 
\hline
\end{tabular}
\vspace{1em}
\normalsize
\caption{Parameters of our dynamical input distributions as
parametrized in (5) referring to an input scale of $Q_0^2\equiv
\mu_{\rm NLO}^2=0.5$ GeV$^2$ at NLO and $Q_0^2\equiv \mu_{\rm LO}^2
= 0.3$ GeV$^2$ at LO.  Since the input gluon distribution turned out
to be insensitive to the polynomial terms in (5), we have set them
to zero ($A_g=B_g=0$).  The total number of degrees of freedom is
$\rm dof = 1739 - 21 = 1718$.  The $\chi^2/\rm dof$ in brackets refers
just to the DIS data where $\rm dof = 1239 - 21 = 1218$.  Furthermore
$\alpha_s(\mu_{\rm NLO}^2)/\pi = 0.1659$ and 
$\alpha_s(\mu_{\rm LO}^2)/\pi = 0.2321$.}
\end{center}
\end{sidewaystable}

\newpage
\begin{table}[th]
\renewcommand{\arraystretch}{1.8} 
\vskip 14pt
\scriptsize
\centering
\begin{center}
\begin{tabular}{|c||c|c|c|c|c|}
\hline
&\multicolumn{5}{|c|}{NLO ($\overline{\rm MS}$)} \\
\hline
& $u_v$ & $d_v$ & $\bar{d} - \bar{u}$ & $\bar{u} + \bar{d}$ & $g$ \\
\hline
N & 0.5889 & 0.2585 & 7.2847 & 0.2295 & 1.3667 \\ 

a & 0.3444 & 0.2951 & 1.2773 & -0.1573 & -0.1050 \\ 

b & 3.7312 & 4.8682 & 18.756 & 8.8819 & 3.3358 \\ 

A & -0.1740 & -1.0552 & -6.3187  & 0.8704 & - \\ 

B & 17.997 & 26.536 & 18.306 & 8.2179 & - \\ 

\hline
$\chi^2/{\rm dof}$ & \multicolumn{5}{|c|}{1.016 (0.955)} \\ 
\hline
$\alpha_s(M_Z^2)$ & \multicolumn{5}{|c|}{0.1178 $\pm$ 0.0021} \\ 
\hline
\end{tabular}
\vspace{1em}
\normalsize
\caption{As Table 1 but for the input parameters in (5) of
the NLO standard fit at an input scale $Q_0^2=2$ GeV$^2$.}
\end{center}
\end{table}

\clearpage
\begin{figure}
\begin{center}
\ifpdf
\includegraphics[width=20.5cm, angle=90]{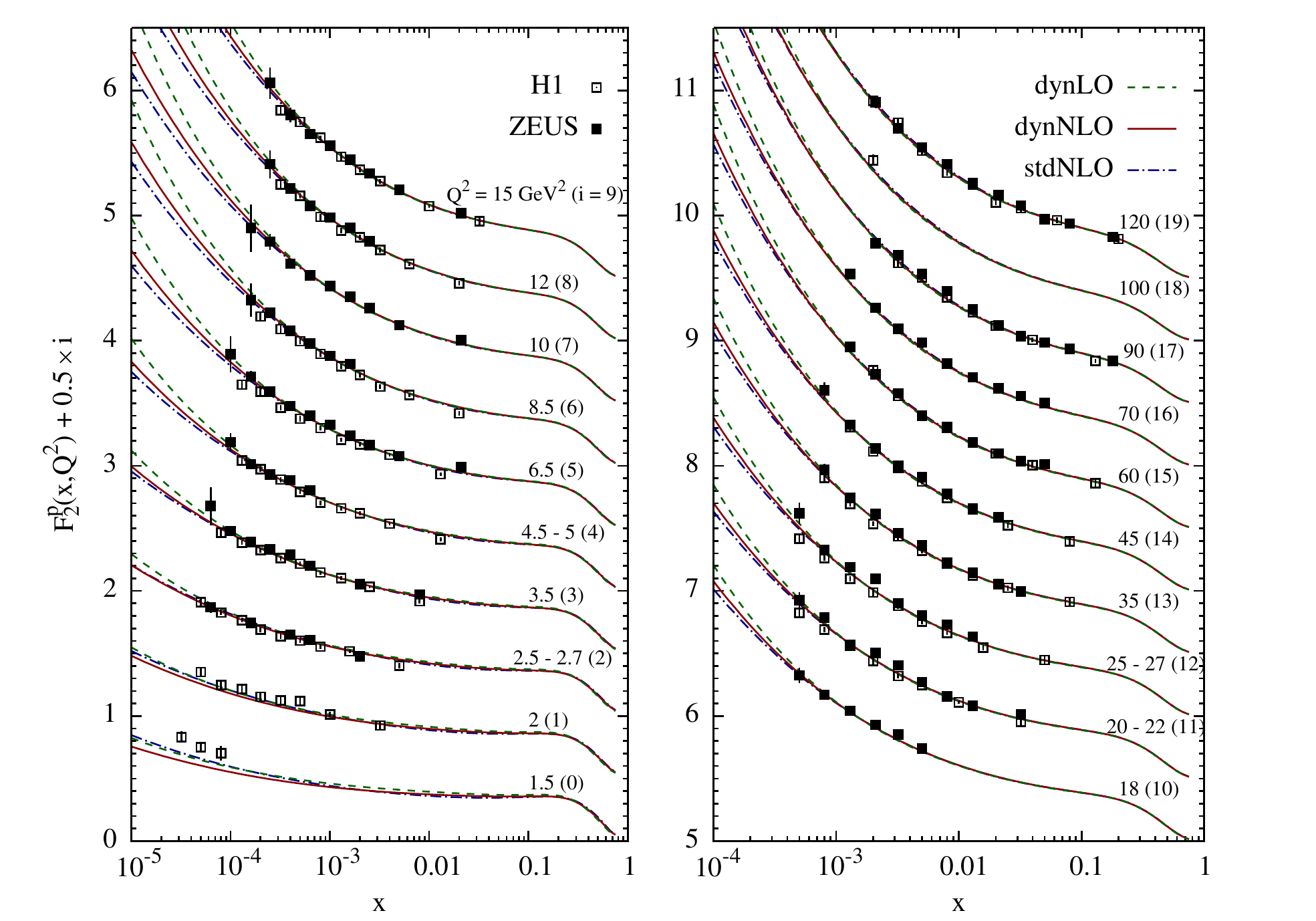}
\else
\includegraphics[width=20.5cm, angle=90]{Fig1_0706.eps}
\fi
\caption{Comparison of our dynamical (dyn) LO and NLO($\overline{\rm MS}$)
as well as standard (std) NLO small--$x$ results for $F_2^p(x,Q^2)$ with
HERA data for $Q^2\geq 1.5$ GeV$^2$ \cite{ref28,ref29}. The parameters of
the valence--like input distributions for the dynamical predictions are
given in Table 1 and the ones for the standard results in Table 2. To
ease the graphical presentation we have plotted $F_2^p(x,Q^2)+0.5\times 
i(Q^2)$ with $i(Q^2)$ indicated in parentheses in the figure for each 
fixed value of $Q^2$.}
\end{center}
\end{figure}

\clearpage
\begin{figure}
\begin{center}
\ifpdf
\includegraphics[width=20.5cm, angle=90]{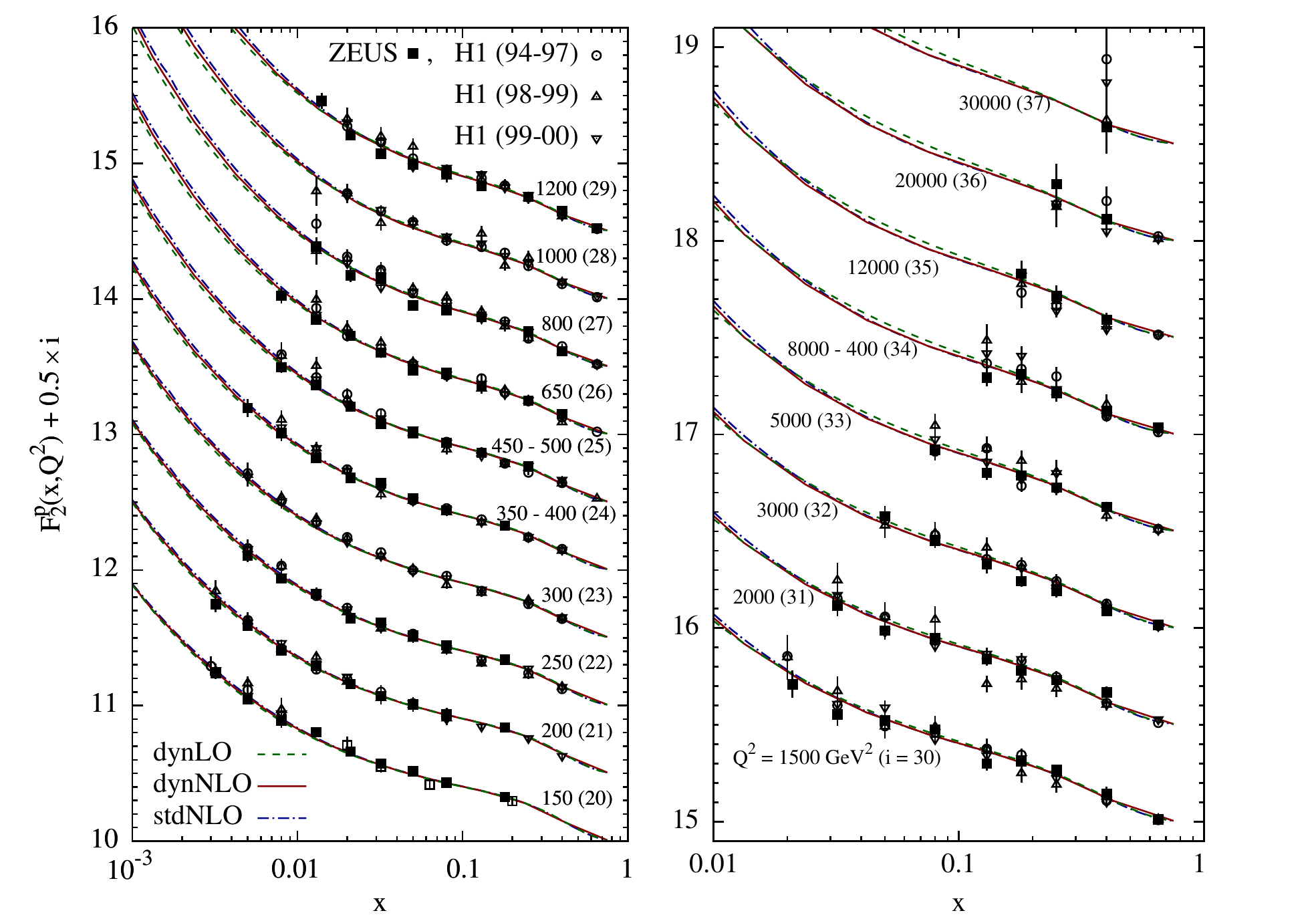}
\else
\includegraphics[width=20.5cm, angle=90]{Fig2_0706.eps}
\fi
\caption{As in Fig.~1 but for large values of $Q^2$ and larger $x$.}
\end{center}
\end{figure}

\clearpage
\begin{figure}
\ifpdf
\includegraphics[width=14.5cm]{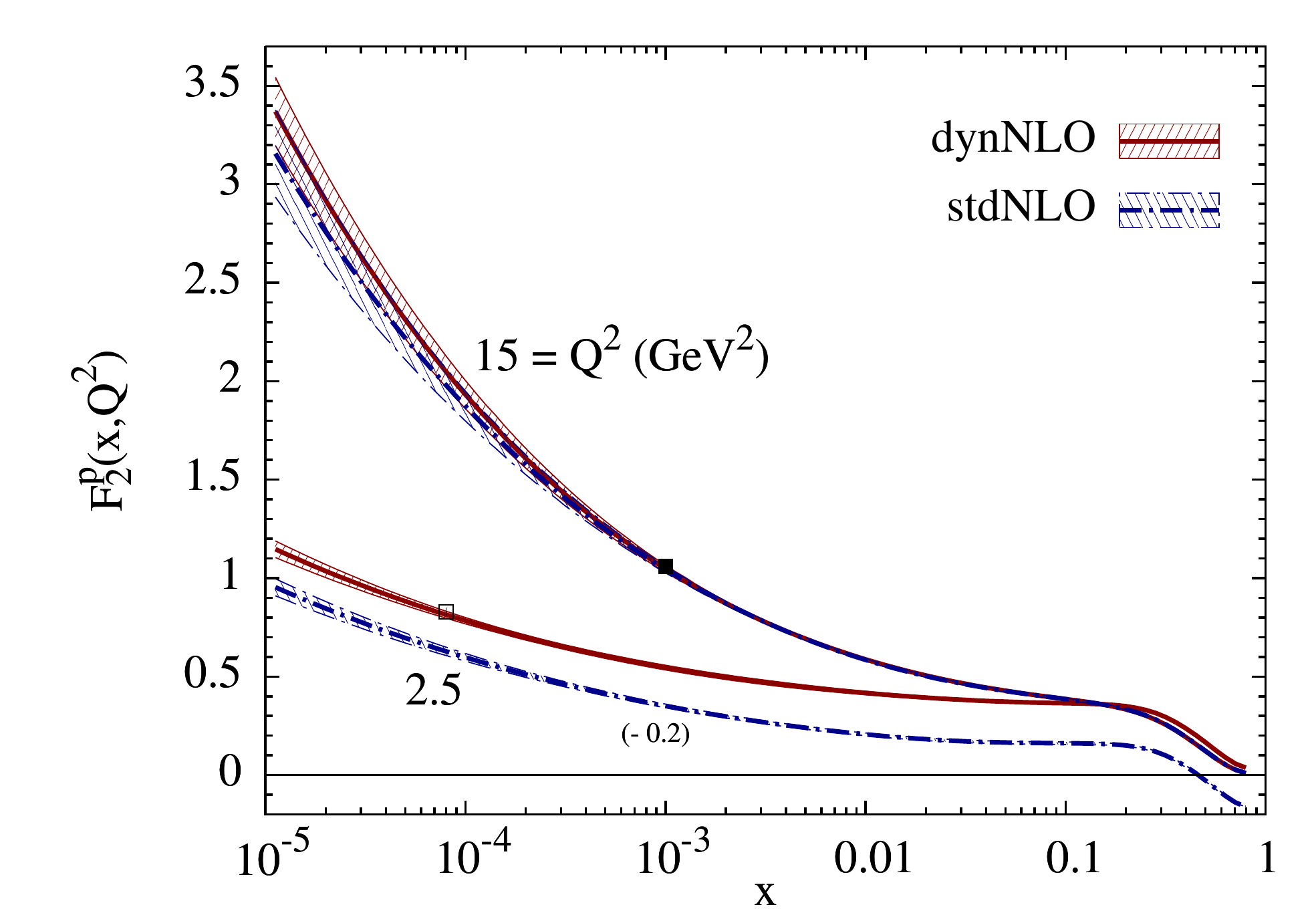}
\else
\includegraphics[width=14.5cm]{Fig3_0706.eps}
\fi
\caption{Typical $\pm 1\sigma$ uncertainty bands of our dynamical and 
standard NLO($\overline{\rm MS})$ results in Fig.~1 for two representative
values of $Q^2$.  To ease the visibility of the two error bands at 
$Q^2=2.5$ GeV$^2$ we have subtracted 0.2 from the stdNLO result as
indicated.  For illustration two H1 and ZEUS data points from Fig.~1 
with their almost invisible errors are shown as well at $Q^2=2.5$ and
15 GeV$^2$, respectively.}
\end{figure}

\clearpage
\begin{figure}
\ifpdf
\includegraphics[width=15.5cm]{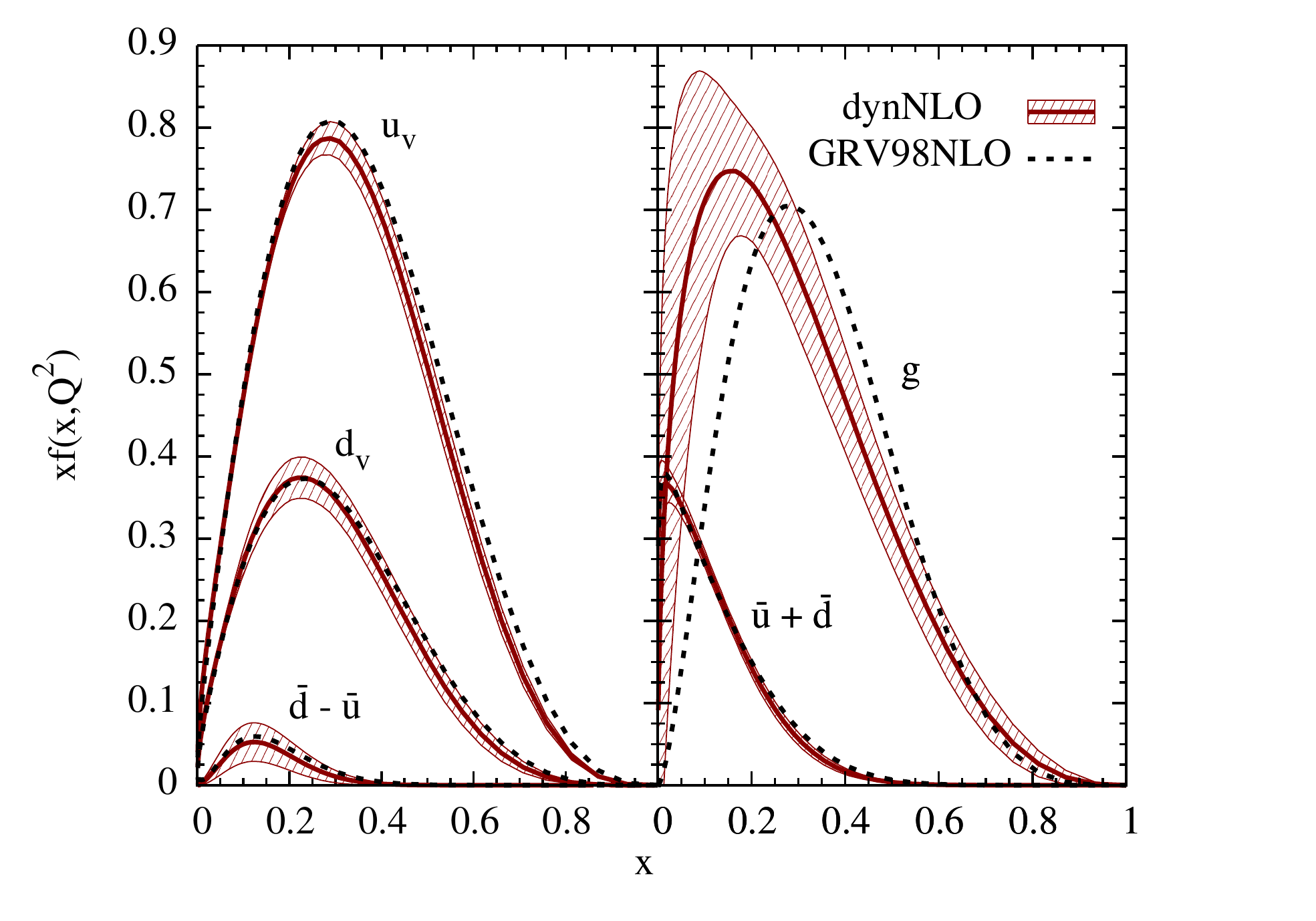}
\else
\includegraphics[width=14.5cm]{Fig4_0706.eps}
\fi
\caption{The valence--like input densities together with their 
$\pm 1\sigma$ uncertainties at $Q_0^2\equiv\mu_{\rm NLO}^2=0.5$ GeV$^2$
for our dynamical NLO($\overline{\rm MS}$) results.  The central curves 
follow from (5) with the 
parameters given in Table 1.  The strange sea $s=\bar{s}$ vanishes at
the input scale.  The NLO GRV98 input \cite{ref8} is also shown by the 
dashed curves for comparison.}
\end{figure}

\clearpage
\begin{figure}
\ifpdf
\includegraphics[width=15.5cm]{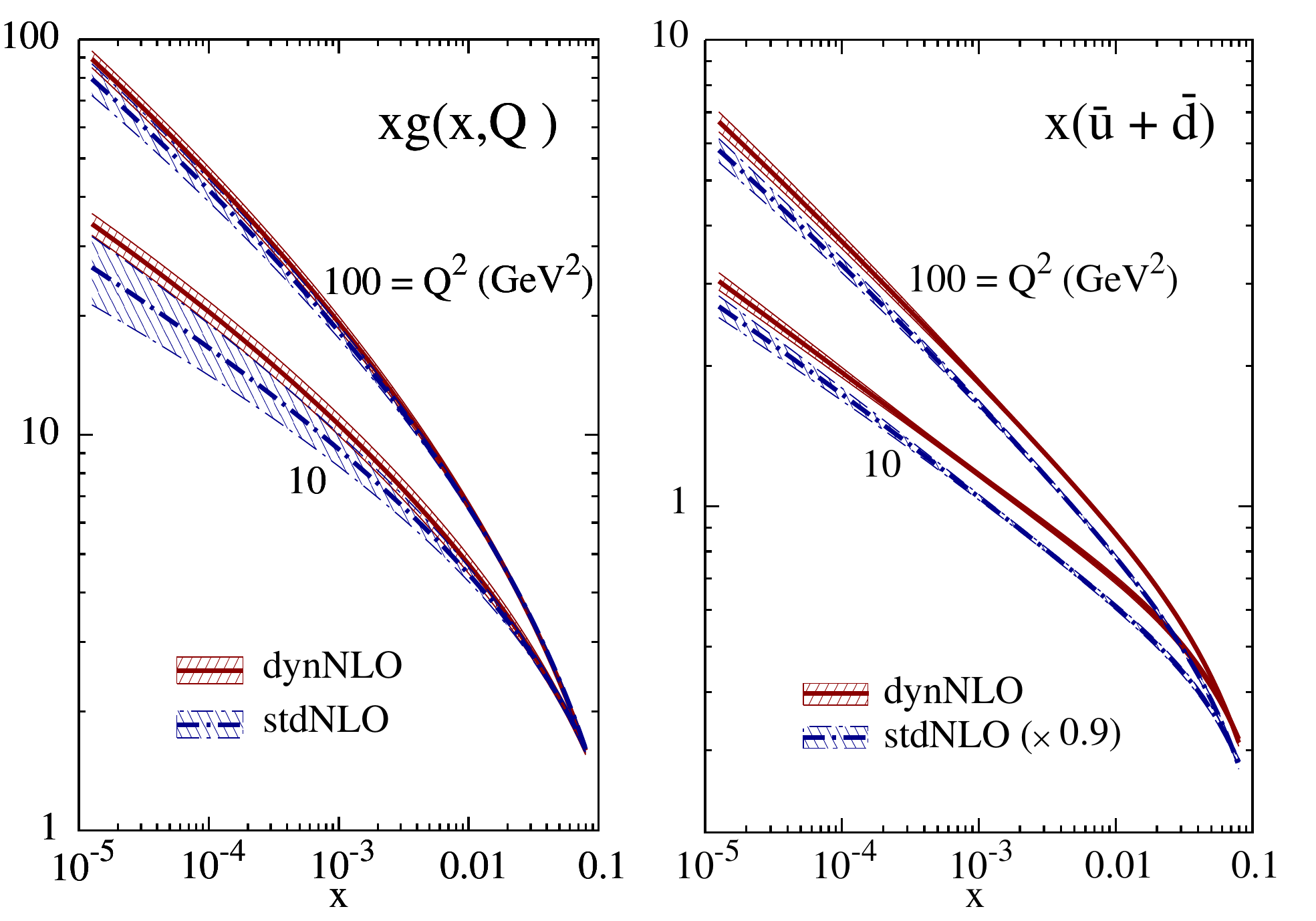}
\else
\includegraphics[width=15.5cm]{Fig5_0706.eps}
\fi
\caption{The small--$x$ NLO($\overline{\rm MS}$) predictions of our
dynamically (radiatively) generated gluon and sea-quark distributions,
together with their $\pm 1\sigma$ uncertainties, as compared to the 
results of a standard fit. To ease the visibility of the two error
bands of $x(\bar{u}+\bar{d})$ we have multiplied the stdNLO results
by 0.9 as indicated. The corresponding GRV98 predictions \cite{ref8}
lie within the $1\sigma$ band of our new dynNLO results.}
\end{figure}

\clearpage
\begin{figure}
\ifpdf
\includegraphics[width=14.5cm]{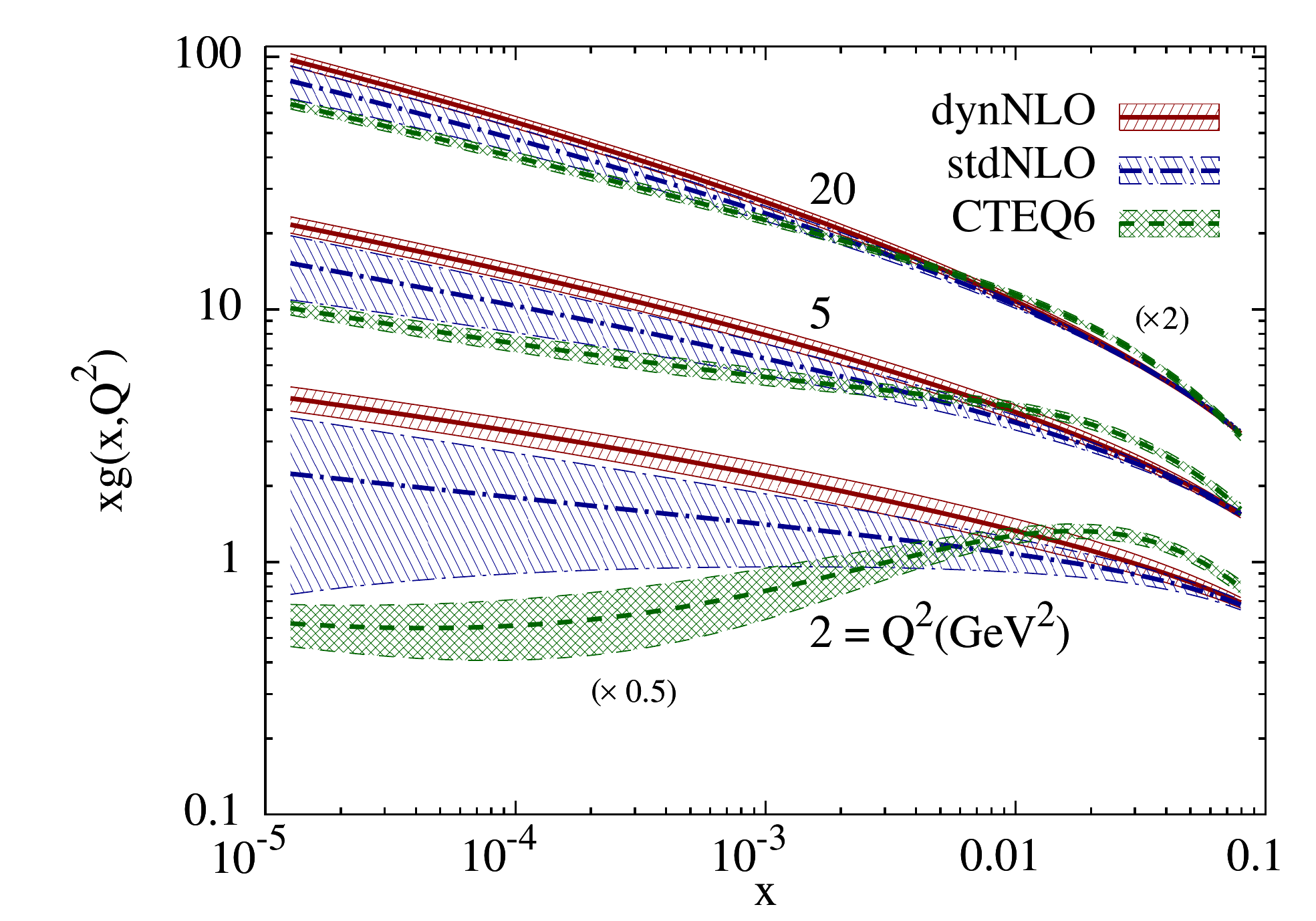}
\else
\includegraphics[width=14.5cm]{Fig6_0706.eps}
\fi
\caption{Comparing the $\pm 1\sigma$ error bands of our dynamical (dyn),
standard (std) and CTEQ \cite{ref2} NLO($\overline{\rm MS}$) gluon
distribution at small $x$ for various fixed values of $Q^2$.  Note that
$Q^2 = 2$ GeV$^2$ is the input scale of the standard fit which is close
to the CTEQ input scale $Q_0^2=m_c^2\simeq 1.7$ GeV$^2$, where the 
standard CTEQ6 fit employs a valence--like gluon input (i.e., $xg(x,Q_0^2)
\to 0$ as $x\to 0$). Due to the sizeably different input scales, the 
CTEQ6 gluon falls up to 30--40\% below our dynNLO gluon for $x<10^{-3}$
and $Q^2>10$ GeV$^2$.  The results at $Q^2=2$ and 20 GeV$^2$ have been
multiplied by 0.5 and 2, respectively, as indicated in the figure.}
\end{figure}

\clearpage
\begin{figure}
\ifpdf
\includegraphics[width=14.5cm]{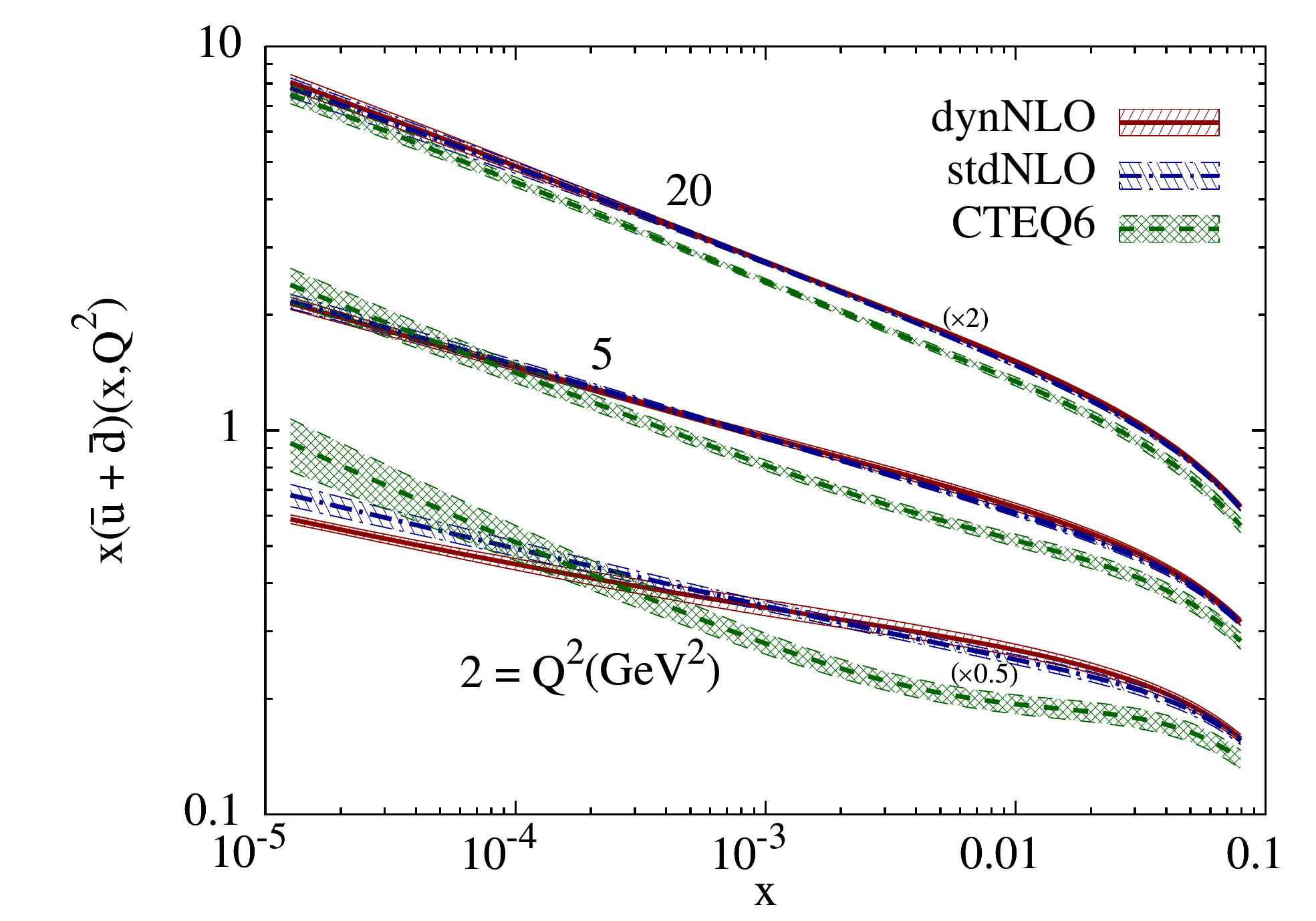}
\else
\includegraphics[width=14.5cm]{Fig7_0706.eps}
\fi
\caption{As in Fig.~6 but for the sea--quark distribution $x(\bar{u}+
\bar{d})$.  Notice that here the CTEQ input distribution at $Q_0^2\simeq
1.7$ GeV$^2$ is not valence--like (i.e., $x(\bar{u}+\bar{d})(x,Q_0^2)
\to \!\!\!\!\!\!\!/ \,\,\,\, 0$ for $x\to 0$).}
\end{figure}

\clearpage
\begin{figure}
\ifpdf
\includegraphics[width=13.5cm]{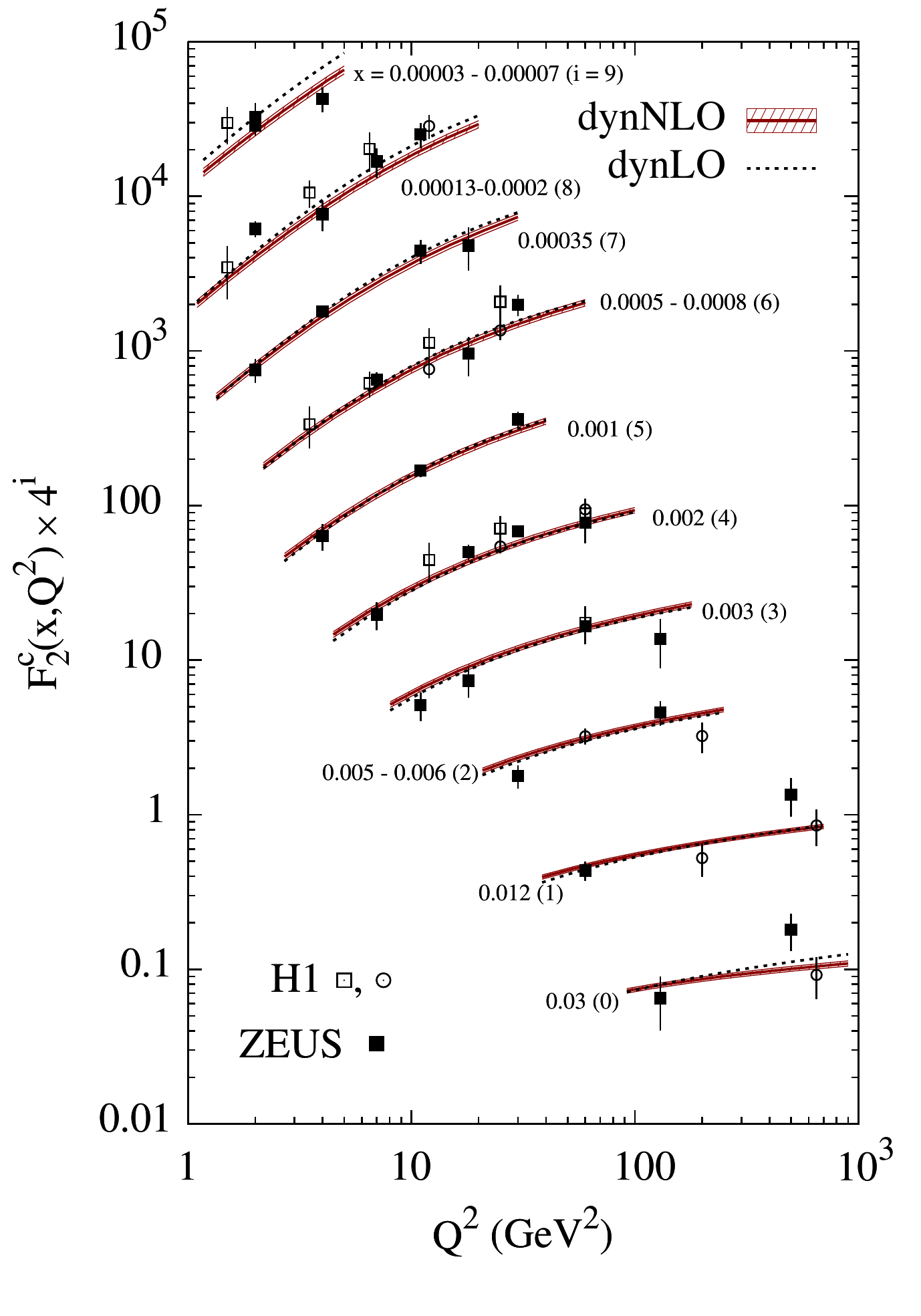}
\else
\includegraphics[width=13cm]{Fig8_0706.eps}
\fi
\caption{The dynamical NLO($\overline{\rm MS}$) predictions for $F_2^c$
in the strict $n_f=3$ FFNS, choosing $\mu_F^2=4m_c^2$ with 
$m_c=1.3$ GeV, together with the $\pm 1\sigma$ uncertainty band. 
For comparison we also display the central LO predictions which are 
entirely due to the $\gamma^*$--gluon fusion subprocess $\gamma^*g\to
c\bar{c}$.  The charm production data as obtained from $D^*$ measurements
are taken from \cite{ref38,ref39} (solid and open squares) and the H1
direct track measurements from \cite{ref40} (open circles).}
\end{figure}

\clearpage
\begin{figure}
\ifpdf
\includegraphics[width=14cm]{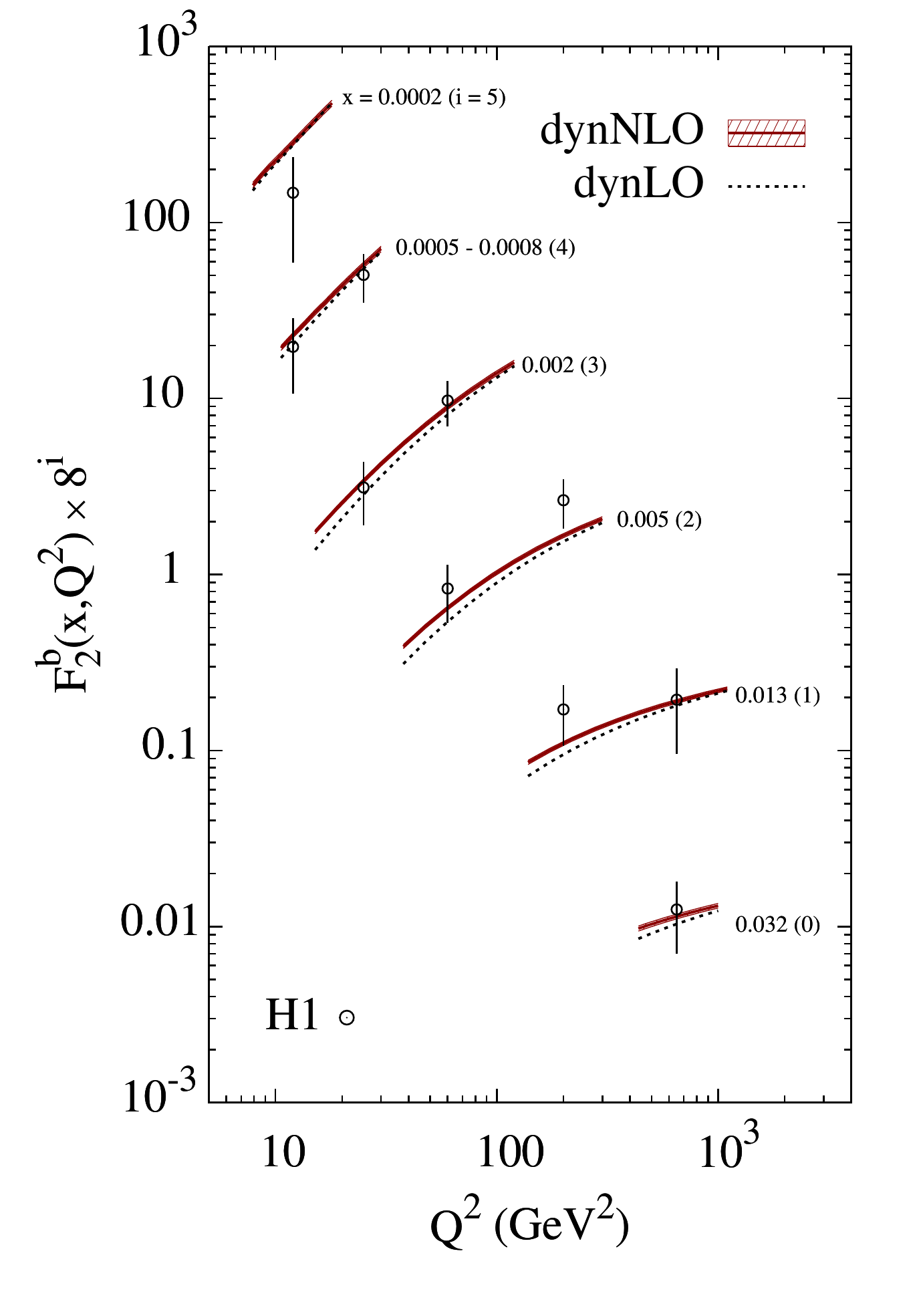}
\else
\includegraphics[width=12.5cm]{Fig9_0706.eps}
\fi
\caption{As in Fig.~8 but for $F_2^b$ with $m_b=4.2$ GeV and the bottom
production data taken from \cite{ref40}.}
\end{figure}

\clearpage
\begin{figure}
\begin{center}
\ifpdf
\includegraphics[width=11cm]{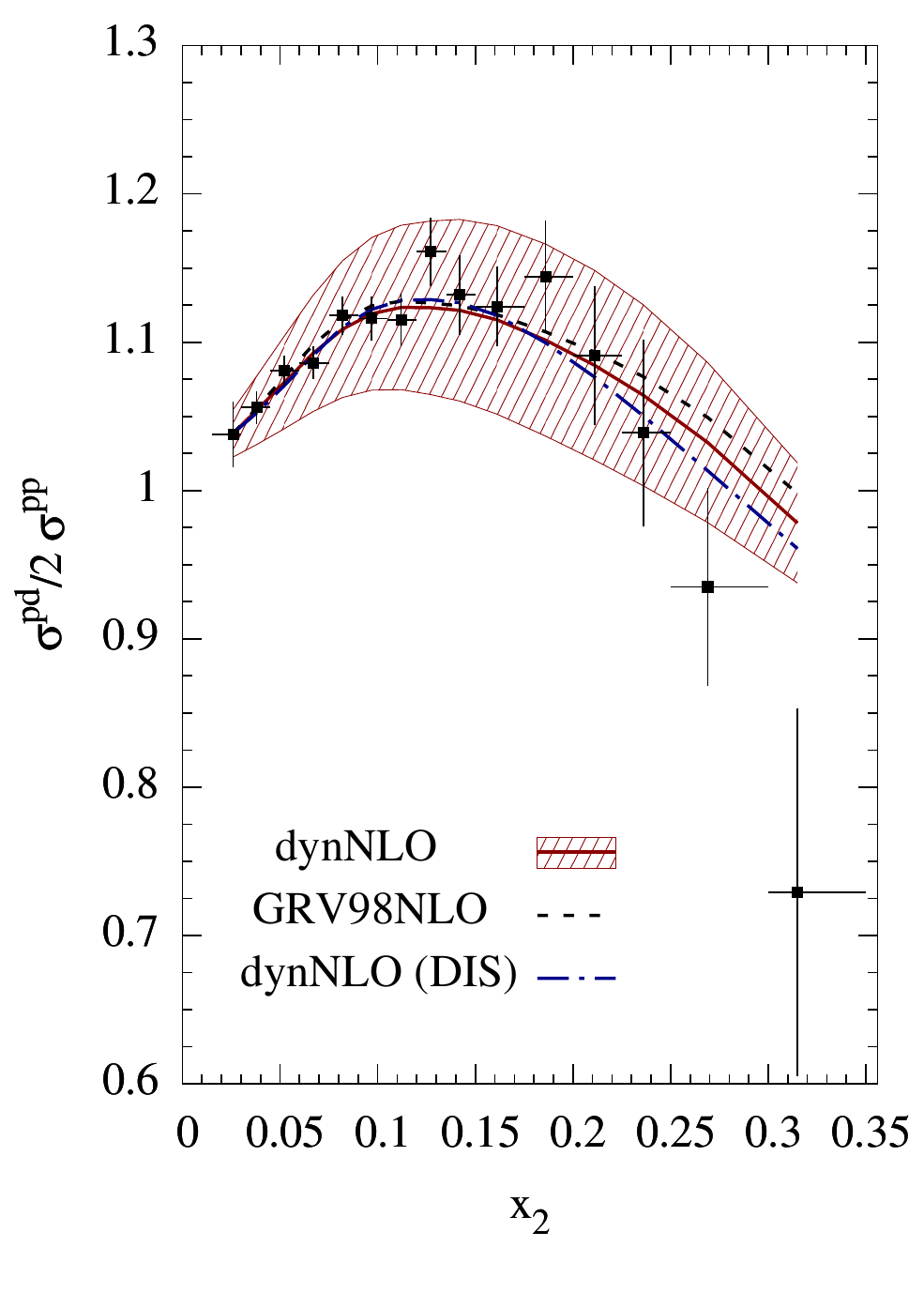}
\else
\includegraphics[width=13cm]{Fig10_0706.eps}
\fi
\end{center}
\caption{Our dynamical NLO result in the $\overline{\rm MS}$
factorization scheme, together with its $\pm 1\sigma$ uncertainty,
for $\sigma^{pd}/2\sigma^{pp}$ appearing in the Drell--Yan asymmetry
$A_{DY} = (\sigma^{pp}-\sigma^{pn})/(\sigma^{pp}+\sigma^{pn})$ as 
a function of the average fractional momentum $x_2$ of the target
partons.  The GRV98 NLO $\overline{\rm MS}$--result \cite{ref8} is
shown for comparison.  The dynamical NLO(DIS) result in the DIS
factorization scheme is shown by the dashed--dotted curve.  The data
for the dimuon mass range $4.6\leq M_{\mu^+\mu^-}\leq 12.9$ GeV are
taken from \cite{ref42}.}
\end{figure}

\clearpage
\begin{figure}
\ifpdf
\includegraphics[width=15cm]{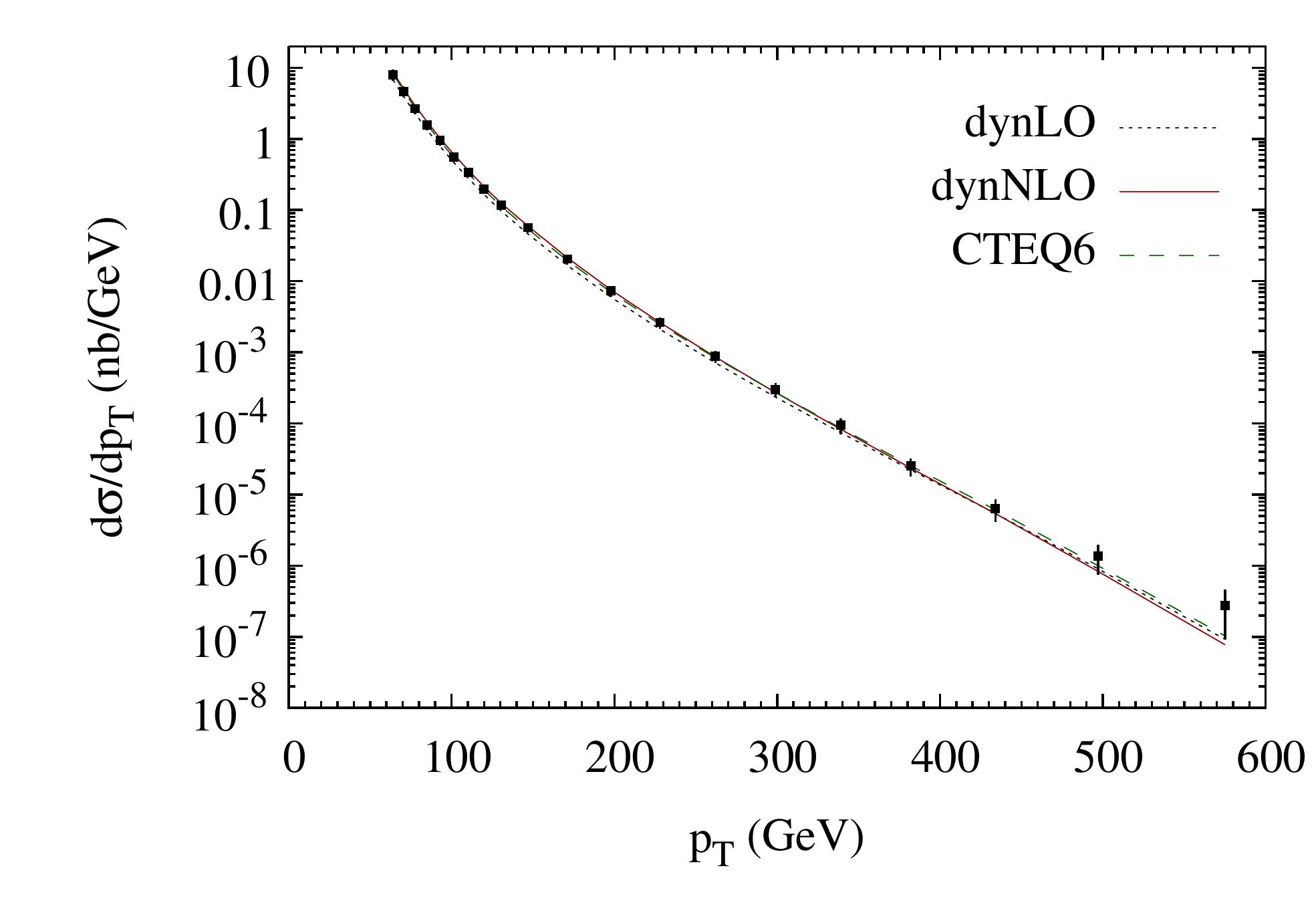}
\else
\includegraphics[width=14.5cm]{Fig11_0706.eps}
\fi
\caption{The $p\bar{p}$ Tevatron high--$p_{\rm T}$ inclusive jet data
\cite{ref46,ref47} compared with our dynamical LO and 
NLO($\overline{\rm MS}$)results, as well as with the NLO CTEQ6 result
\cite{ref2}.}
\end{figure}

\clearpage
\begin{figure}
\ifpdf
\includegraphics[width=14.5cm]{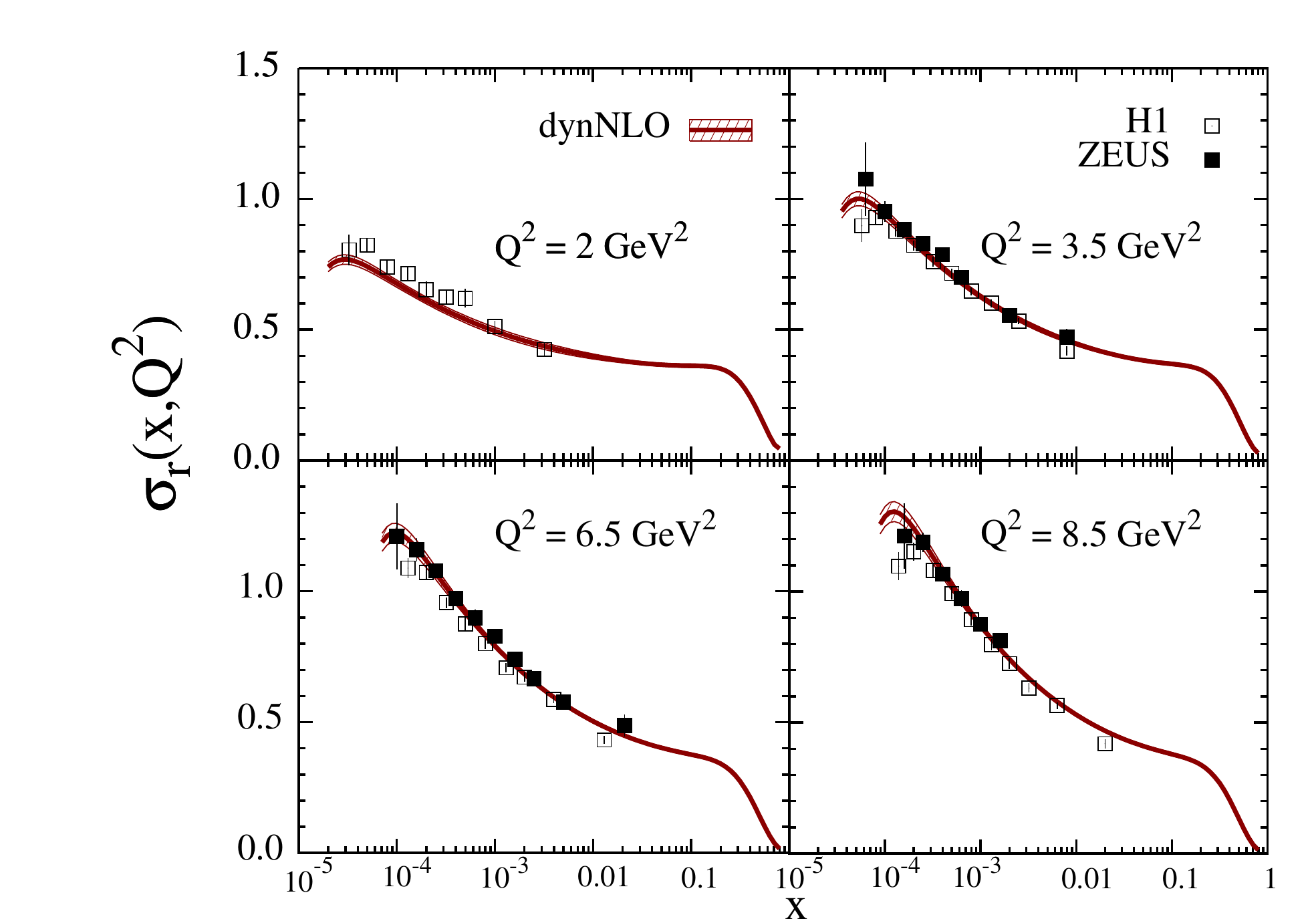}
\else
\includegraphics[width=14.5cm]{Fig12_0706.eps}
\fi
\caption{The dynamical NLO($\overline{\rm MS}$) predictions, together
with their $\pm 1\sigma$ uncertainties, for the `reduced' DIS cross
section $\sigma_r(x,Q^2)=F_2-(y^2/Y_+)F_{\rm L}$.  The HERA data for some
representative fixed values of $Q^2$ are from \cite{ref28,ref29}.}
\end{figure}

\clearpage
\begin{figure}
\ifpdf
\includegraphics[width=14.5cm]{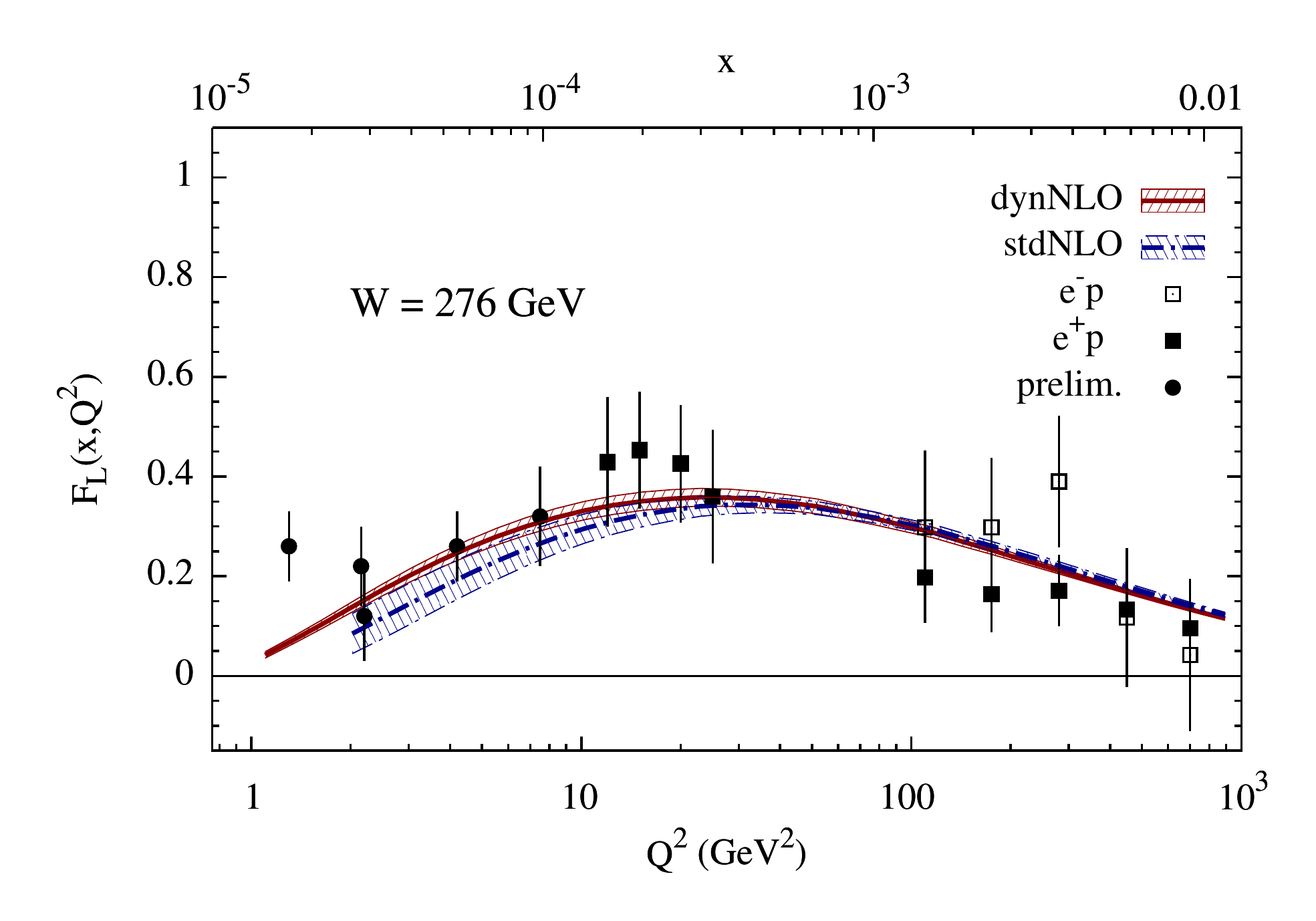}
\else
\includegraphics[width=14.5cm]{Fig13_0706.eps}
\fi
\caption{Dynamical and standard NLO($\overline{\rm MS}$) results for
$F_{\rm L}(x,Q^2)$ together with their $\pm 1\sigma$ uncertainty bands.
The (partly preliminary) H1 data \cite{ref28,ref64} are at fixed
$W\simeq 276$ GeV.}
\end{figure}

\clearpage
\begin{figure}
\ifpdf
\includegraphics[width=14cm]{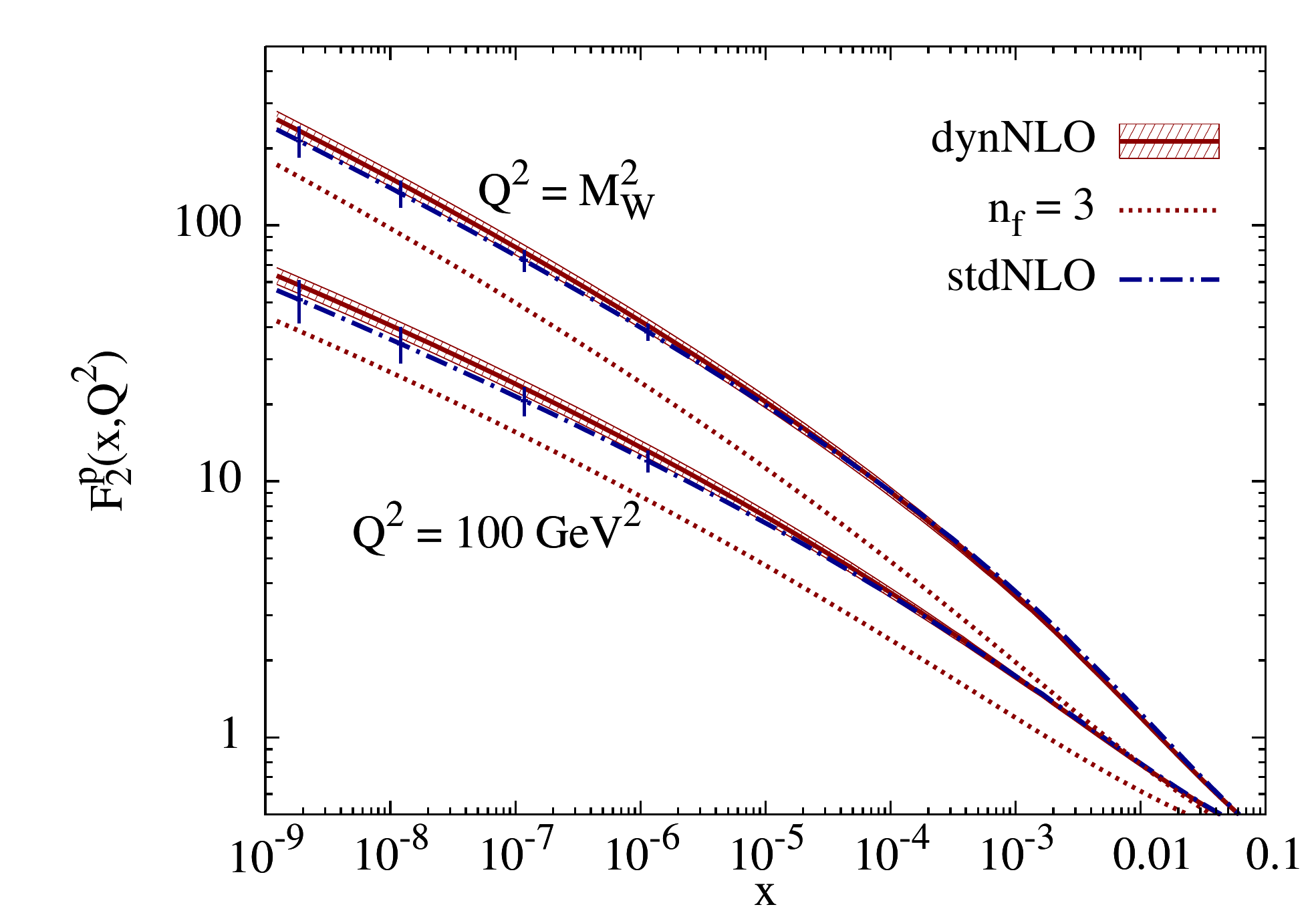}
\else
\includegraphics[width=14cm]{Fig14_0706.eps}
\fi
\caption{Dynamical NLO($\overline{\rm MS}$) predictions for 
$F_2^p(x,Q^2)$, together with their $1\sigma$ uncertainties, for 
extremely small values of $x$.  The $1\sigma$ uncertainties of the 
standard (std) NLO fit extrapolations are shown by the vertical bars.
The dotted curves are the contributions from the light ($n_f=3$)
quarks and gluons to $F_2^p$ for the dynamical (dyn) NLO result. In
other words, the difference between the dotted and solid curves is
due to NLO heavy quark (charm, bottom) contributions which derive 
from photon--gluon (quark) fusion processes.  The dynamical GRV98
predictions \cite{ref8} lie within the $\pm 1\sigma$ band of our
present dynNLO predictions.}
\end{figure}

\clearpage
\begin{figure}
\ifpdf
\includegraphics[width=14.5cm]{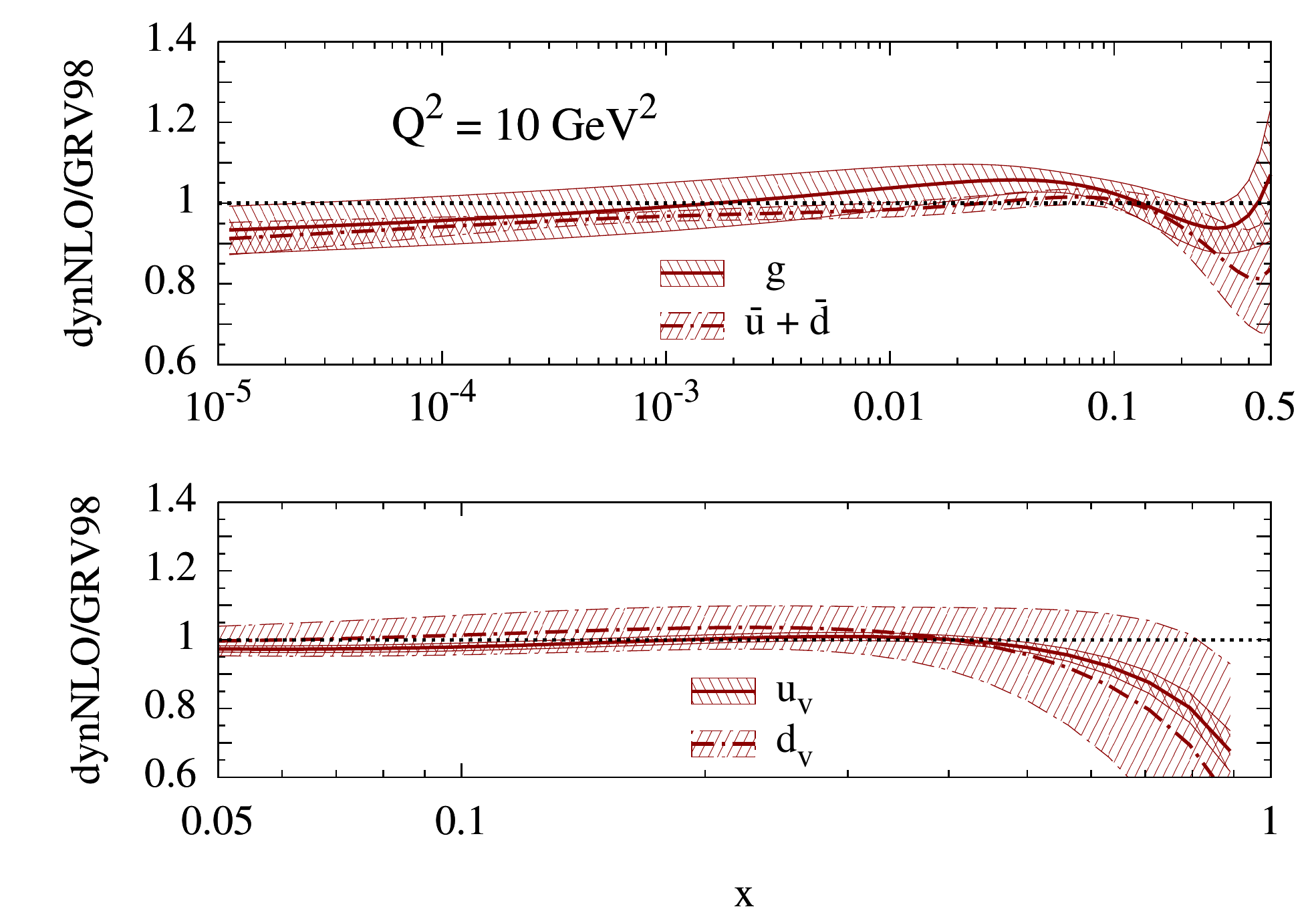}
\else
\includegraphics[width=14.5cm]{Fig15_0706.eps}
\fi
\caption{Comparing the present dynamical dynNLO($\overline{\rm MS}$)
parton distributions with the previous ones of GRV98 \cite{ref8} at
$Q^2=10$ GeV$^2$.}
\end{figure}

\clearpage
\begin{figure}
\ifpdf
\includegraphics[width=14.5cm]{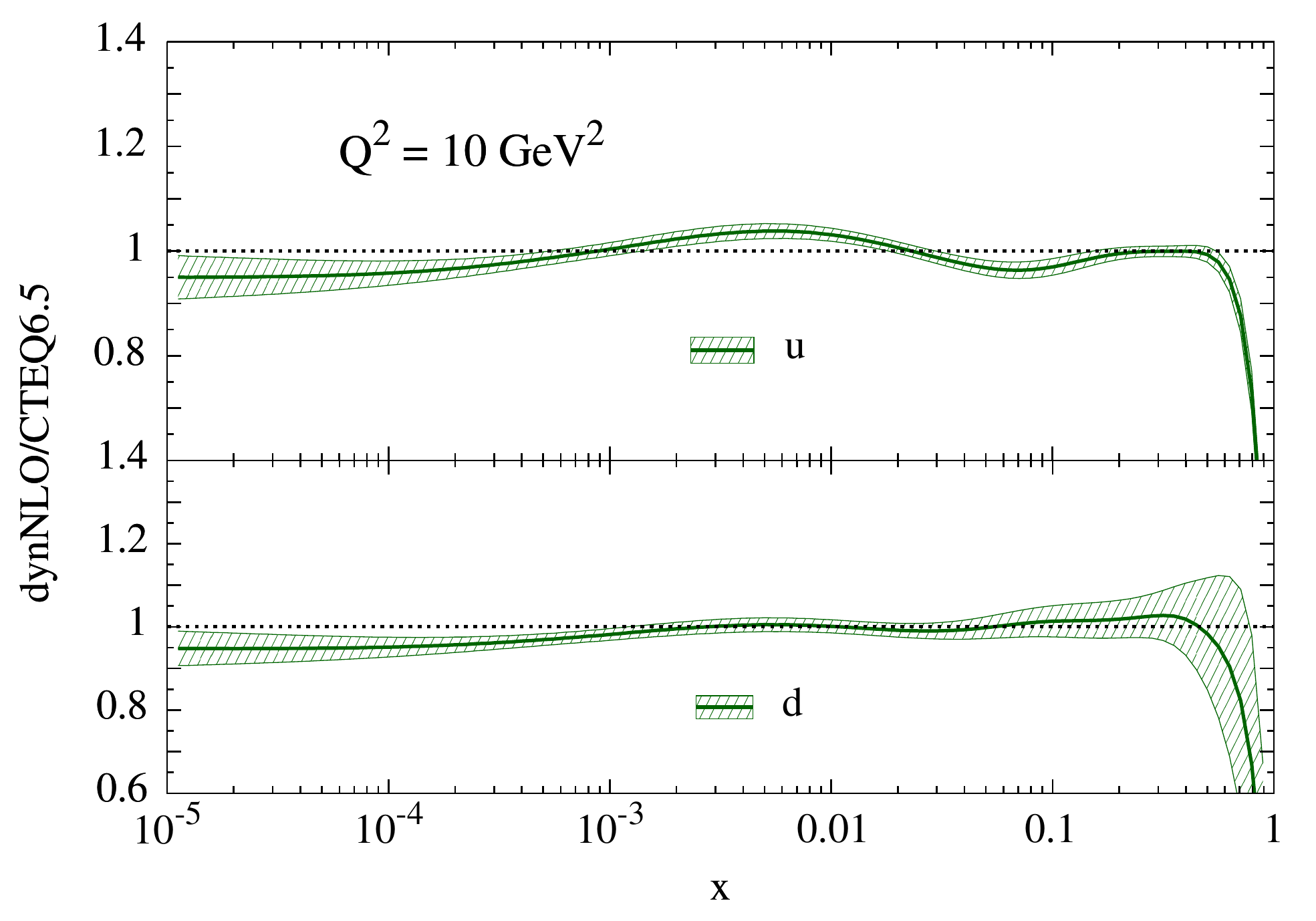}
\else
\includegraphics[width=14.5cm]{Fig16_0706.eps}
\fi
\caption{Comparing the present dynamical NLO($\overline{\rm MS}$)
$u=u_v+\bar{u}$ and $d=d_v+\bar{d}$ distributions with the ones of
CTEQ6.5 \cite{ref70} at $Q^2=10$ GeV$^2$.  These ratios remain
practically unchanged at higher scales, like $Q^2=M_W^2$ relevant
for $W^{\pm}$ production.  The shaded areas represent the estimated
$\pm 1\sigma$ uncertainty band of our dynNLO analysis.  Notice that
in the relevant small--$x$ region these ratios would be practically
unaltered if the GRV98 distributions \cite{ref8} were used instead 
of the dynNLO ones, since dynNLO/GRV98 $\simeq 1$ as evident from
Fig.~15.}
\end{figure}

\end{document}